\documentclass{article}
\usepackage{graphicx}
\usepackage{palatino}
\usepackage{framed}
\usepackage{amsmath}
\usepackage{amsfonts}
\usepackage{amssymb}
\usepackage[T1]{fontenc}
\usepackage{soul}
\usepackage{float}
\usepackage[hidelinks, breaklinks]{hyperref} 
\usepackage[hyphenbreaks]{breakurl}

\usepackage{multirow} 
\usepackage{subcaption}
\usepackage{authblk}

\usepackage[round]{natbib}
\usepackage{pb-diagram} 
\usepackage{ulem}

\usepackage{anysize}
\marginsize{2cm}{2.5cm}{2.5cm}{2cm}

\usepackage[textwidth=3cm, textsize=footnotesize]{todonotes} 

\newcommand{\cB}{{\cal B}}

\author[1]{Sergios Agapiou}
\author[1]{Andreas Anastasiou}
\author[1]{Anastassia Baxevani}
\author[1]{Tasos Christofides}
\author[2]{Elisavet Constantinou}
\author[3]{Georgios Hadjigeorgiou}
\author[4]{Christos Nicolaides}
\author[3]{Georgios Nikolopoulos}
\author[1]{Konstantinos Fokianos}

\affil[1]{Department of Mathematics \& Statistics, University of Cyprus}
\affil[2]{Cyprus Ministry of Health}
\affil[3]{Medical School, University of Cyprus}
\affil[4]{Department of Business \& Public Administration, University of Cyprus}
{
    \makeatletter
    \renewcommand\AB@affilsepx{: \protect\Affilfont}
    \makeatother

    \affil[ ]{Email}

    \makeatletter
    \renewcommand\AB@affilsepx{, \protect\Affilfont}
    \makeatother

    \affil[1,3,4]{{agapiou.sergios,anastasiou.andreas,baxevani,
    tasos, hadjigeorgiou.georgios, nicolaides.christos, nikolopoulos.georgios,fokianos}@ucy.ac.cy}
    \affil[2]{EConstantinou@moh.gov.cy}
}

\title{Modeling of Covid-19 Pandemic in Cyprus}
\date{\today}

\begin{document}
\maketitle

\begin{abstract}
\noindent

The Republic of Cyprus is a small island in the southeast of Europe and member of the European Union. The first wave of COVID-19 in Cyprus started in early March, 2020 (imported cases) and peaked in late March-early April. The health authorities responded rapidly and rigorously to the COVID-19 pandemic by scaling-up testing, increasing efforts to trace and isolate contacts of cases, and implementing measures such as closures of educational institutions, and travel and movement restrictions. The pandemic was also a unique opportunity that brought together experts from various disciplines including epidemiologists, clinicians, mathematicians, and statisticians. The aim of this paper is to present the efforts of this new, multidisciplinary research team in modelling the COVID-19 pandemic in the Republic of Cyprus.


\bigskip


\pagebreak
\end{abstract}

\section{Introduction}

Coronavirus disease 2019 (COVID-19), an infection caused by the novel coronavirus SARS-CoV-2  (\cite{Coronastudy(2020)}) that first emerged in Wuhan, China, (\cite{Zhuetal(2020)}), counts now more than 25 million cases and has claimed nearly 850,000 lives (\cite{WHOreport197(2020)}). Despite some advances in therapy (\cite{Beigeletal(2020)}) and considerable progress in vaccine development, with some vaccine candidates reaching phase III trials (\cite{Jacksonetal(2020)}), there are still many gaps in our understanding of the new pandemic disease including some epidemiological parameters.
Epidemic modelling is a fundamental component of epidemiology, especially with regards to infectious diseases. Following the pioneering work of R. Ross, W. Kermack, and McKendrick in early twentieth century (\cite{KermackandMcKendick(1927)}), the discipline has established itself and comprises a major source of information for decision makers. For instance, in the United Kingdom, the Scientific Advisory Group of Emergencies (SAGE) is a major body that collects evidence from multiple sources including inputs from mathematical modelling to advice the British government on its response to the complex COVID-19 situation; for more information see
\href{https://www.gov.uk/government/collections/scientific-evidence-supporting-the-government-response-to-coronavirus-covid-19}{\underline{this link}}.

In the context of the COVID-19 pandemic, expert opinions can help decision makers comprehend the status of the pandemic by collecting, analyzing, and interpreting relevant data and developing scientifically sound methods and models. An exact model that would describe perfectly the data is usually not feasible and of limited scope; hence scientists usually aim for models that allow a statistical simulation of synthetic data. At the same time, models can also approximate the dynamics of the disease and discover important patterns in the data. In this way, researchers can study various scenarios and understand the likely consequences of government interventions. Finally, the proposed models could motivate the conduct of further studies about the evolution of both infectious and non-infectious diseases of public interest.

Here we report our work including results from statistical and mathematical models used to understand the epidemiology of COVID-19 in Cyprus,  during the time period starting from the beginning of March till the end of May 2020. We propose a range of different models that  capture different aspects of the COVID-19 pandemic. The analysis consists of several methods applied to understand the evolution of pandemics in the long and short run. We use change-point detection,  count time series methods and compartmental models for short and long term projections, respectively.
We estimate the effective reproduction number by using three different methods and obtain consistent results irrespective of the
method used. Results are cross-validated against observed data with considerable consistency.  Besides providing a comprehensive data analysis we illustrate the importance of mathematical models to epidemiology.

\section{Statistical Methods}
In this section, after a brief introduction to the testing protocol, we introduce the different techniques and models that have  been used for the modelling and analysis of the COVID-19 infections in Cyprus.

\subsection{Cyprus COVID-19 surveillance system}

The Unit for Surveillance and Control of Communicable Diseases (USCCD) of the Ministry of Health operates COVID-19 surveillance. The lab-based surveillance system consists of 19 laboratories (7 public and 12 private) that carry out molecular diagnostic testing for SARS-CoV-2. Sociodemographic, epidemiological, and clinical data of individuals with SARS-CoV-2 infection are routinely collected from laboratories and
clinics, and reported to an electronic platform of the USCCD.
A confirmed COVID-19 case is a person, symptomatic or asymptomatic, with a respiratory swab (nasopharynx and/or pharynx) positive for SARS-CoV-2 by a real-time reverse-transcription polymerase chain (rRT-PCR) assay. Cases are considered imported if they have travel history from an affected area within 14 days of the disease onset. Locally-acquired cases are individuals who test positive for SARS-CoV-2 and have the earliest onset date in Cyprus without travel history from affected areas. People with symptomatic COVID-19 are considered recovered after the resolution of symptoms and two negative tests for SARS-CoV-2 at least 24-hour apart from each other. For asymptomatic cases, the negative tests to document virus clearance are obtained at least 14 days after the initial positive test. A person with a positive test at 14 days is further isolated for one week and finally released at 21 days after the initial diagnosis without further laboratory tests.
Testing approaches in the Republic of Cyprus included: a) targeted testing of suspect cases and their contacts; of repatriates at the airport and during their 14-day quarantine; of teachers and students when schools re-opened in mid-May; of employees in essential services that continued their operation throughout the first pandemic wave (e.g., customer services, public domain); and of health-care workers in public hospitals, and b) population screenings following random sampling in the general population of most districts and in two municipalities with increased disease burden. By June 2nd 2020, 120,298 PCR tests had been performed (13,734.2 per 100,000 population).
Public health measures were taken in 4 phases: Period 1 (10 - 14 March, 2020) included closures of educational institutions and cancellation of public gatherings (>75 persons); Period 2 (15 - 23 March, 2020) involved closure of entertainment areas (for instance, malls, theatres, etc), allowance of 1 person per 8 square meters in public service areas, and restrictions to international travel (for example, access to the Republic of Cyprus was permitted only for specific persons and after SARS-CoV-2 testing); Period 3 (24 - 30 March, 2020) included closure of most retail services; and Period 4 (31 March - 3 May) included the suspension of incoming flights with few exceptions (for instance, repatriated Cypriot citizens), stay at home order, and night curfew.

\subsection{Change-point Analysis and Projections}
\label{subsub:cpt_detection_methods}

{\raggedright{Change-point detection is an active area of statistical research that has attracted a lot of interest}} in recent years and plays an essential role in the development of the mathematical sciences. A non-exhaustive list of application areas includes financial econometrics  \citep{Schroeder_Fryzlewicz}, credit scoring \citep{Bolton_Hand}, and bioinformatics \citep{Olshen_Venkatraman}. The focus is on the so-called \textit{a posteriori} change-point detection, where the aim is to estimate the number and locations of certain changes in the behaviour of a given data sequence. For a review of methods of inference for single and multiple change-points (especially in the context  of time series) under the a-posteriori framework, see \cite{Jandhyala_review}. The aim is to estimate the number and locations of certain changes in a stream of data. Detecting these change-points enables us to separate the data sequence into homogeneous segments, leading to a more flexible modeling approach. Advantages of discovering such heterogeneous segments include
interpretation and forecasting. Interpretation naturally associates the detected change-points to  real-life events or/and political decisions. In this way,
a better description   of the observed process and the impact of any intervention can be communicated.  Forecasting, is based on  the final detected segment which is  important as it allows for more accurate prediction of future values of the data sequence at hand. Methods developed in this context are based on a given model.
For the purpose of this  paper, we work with the following signal-plus-noise model
\begin{equation}
\label{our_model1}
X_t = f_t + \sigma\epsilon_t, \quad t=1,2,\ldots,T,
\end{equation}
where $X_t$ denotes the daily incidence COVID-19 cases
and $f_t$ is a deterministic signal with structural changes at certain time points.  Details about $f_{t}$ are given below.  The sequence  $\epsilon_t$ consists of independent and identically distributed (iid) data  with mean zero and variance equal to one  and $\sigma >0$.
We denote the number of change-points by $K$ and their respective locations by
$r_1, r_2, \ldots, r_K$. The locations  are unknown and the  aim is to estimate them based on \eqref{our_model1}.
The daily incidence cases  of the COVID-19 outbreak in Cyprus is investigated by using the following two models for $f_{t}$ of \eqref{our_model1}:
\begin{enumerate}
    \item {\textbf{Continuous, piecewise-linear signals:}} $f_{t} = \mu_{j,1} + \mu_{j,2}t$, for $t = r_{j-1} + 1, r_{j-1}+2, \ldots, r_{j}$ with the additional constraint of $\mu_{k,1} + \mu_{k,2}r_{k} = \mu_{k+1,1} + \mu_{k+1,2}r_{k}$ for $k=1,2,\ldots,N$. The change-points, $r_k$, satisfy $f_{r_k-1} + f_{r_k+1}\neq 2f_{r_k}$.
    \item \textbf{Piecewise-constant signals:} $f_t = \mu_j$ for $t = r_{j-1}+1,r_{j-1}+2,\ldots,r_j$, and $f_{r_j}\neq f_{r_j+1}.$
\end{enumerate}
In this work, we are using the Isolate-Detect (ID) methodology of \cite{Anastasiou_Fryzlewicz} to detect changes based on \eqref{our_model1} by using
linear and constant signals, as described above; see Appendix \ref{sec:ISDmethod} for a description of the method.

\subsection{Count Time Series Methodology}

The  analysis of count time series data (like daily incidence data we consider in this work)
has attracted considerable  attention, see
\citet[Sec 4 \& 5]{KedemandFokianos(2002)} for several references and  \citet{Fokianos(2015)} for a more recent review
of  this research  area.
In what follows,  we take the point of view of generalized linear
modelling as advanced by \citet{McCullaghandNelder(1989)}.
This framework naturally  generalizes  the traditional ARMA methodology  and includes several complicated data generating processes besides count data such as  binary and categorical data. In addition, fitting of such models can be carried out by likelihood methods; therefore testing, diagnostics and all type
of likelihood arguments are available to the data analyst.

The logarithmic function is the most popular
link function for  modeling  count data. In fact, this choice
corresponds to the canonical link  of generalized linear models.
Suppose that  $\{X_t \}$ denotes  a daily incidence time series  and
assume, that given the past, $X_{t}$ is conditionally Poisson distributed with
mean $\lambda_{t}$. Define  $\nu_{t} \equiv \log \lambda_{t}$.
A log-linear model with feedback  for the analysis of count time series (\citet{FokianosandTjostheim(2011)}) is defined as
\begin{equation}
\nu_{t}  =   d+a_{1} \nu_{t-1} + b_{1} \log(X_{t-1}+1).
\label{eq:log-linear model}
\end{equation}
In general, the    parameters $d,a_{1},b_{1}$  can be positive or negative but they need to satisfy certain conditions to obtain stability of the model. The inclusion of the hidden process makes the mean of the process to depend on the long-term past values of the observed data.
Further discussion on model \eqref{eq:log-linear model} can be found
in  Appendix \ref{Sec:CountTimeSeries}
which also includes some discussion
about interventions. An intervention is an unusual event that has a temporary or a permanent impact on the observed process. Computational methods for discovering interventions, in the context of \eqref{eq:log-linear model}, under a general mixed Poisson framework have been discussed by \cite{tscountR}.
In this work, we will consider additive outliers (AO) defined by
\begin{equation}
 \nu_{t}  =   d+a_{1} \nu_{t-1} + b_{1} \log(X_{t-1}+1) + \sum_{k=1}^{K}\gamma_{k} I(t= r_{k})
\label{eq:loglinAo}
\end{equation}
where the notation follows closely that
of Sec. \ref{subsub:cpt_detection_methods} and  $I(.)$ denotes the indicator function.
Inclusion of the indicator function
shows that at the time point $r_{k}$, the mean process has a temporary shift whose effect is measured
by the parameter $\gamma_{k}$ but in the log-scale. Other type of interventions can be  included (see Appendix \ref{Sec:CountTimeSeries}) whose effect
can be permanent and, in this sense, intervention analysis and change-point detection methodologies address similar problems but from a different point
of view.
Model fitting  is based on maximum likelihood estimation and
its implementation has been described in detail by \citet{tscountR}.

\subsection{Compartmental Models}


Compartmental models in epidemiology, like the Susceptible-Infectious-Recovered (SIR) and Susceptible-Exposed-Infectious-Recovered (SEIR)
models and their   modifications, have been used to model infectious diseases since the early 1920’s (see \cite{KeelingandRohane(2008)}, \cite{nicolaides2020hand} among others).
The basic assumptions for these models are the existence of a  closed community, i.e without influx of new susceptibles or mortality due to other causes, with a fixed population, say $N$, and also  that the individuals who recover from the illness are immune and do not become susceptible again. In the basic SEIR model, at any point in time  $t$, each individual is either   susceptible ($S(t)$), exposed ($E(t)$), infectious ($I(t)$) or recovered ($R(t)$, including death). The epidemic starts at time $t=0$ with one infectious individual, usually thought of as being externally infected, and the rest of the population being susceptible.  People progress between the different compartments and this motion is described usually through a system of ordinary differential equations that can be put in a stochastic framework.

A variety of SEIR modifications and extensions exist in the literature, and a multitude of them emerged recently because of  the COVID-19 epidemic. In this work, we consider four such modifications, based on the models proposed in  \cite{bib:Peng2020} and \cite{Li2020} for the analysis of the COVID-19 epidemic in Wuhan and the rest of the Chinese provinces.   

\subsubsection{Compartmental Model 1}
\label{sec:compmodel1}

\begin{figure}[H]
	\centering
\includegraphics[width=.6\textwidth]{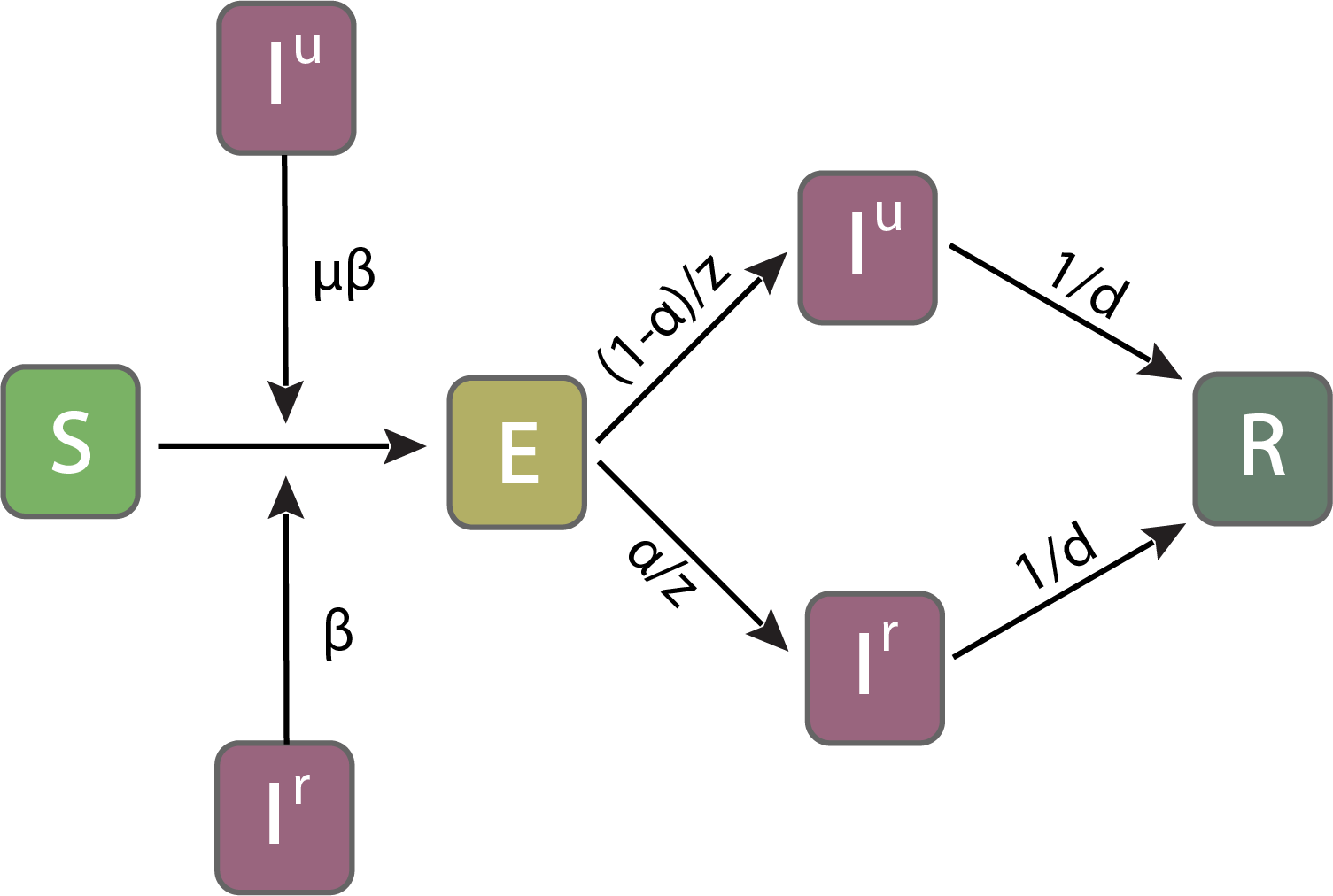}
  \caption{Representation of the SEIR model as described by \cite{Li2020}.}
  \label{fig:seir_li_et_al}
\end{figure}

Initially, we employ the SEIR model  based on the meta-population
model of \cite{Li2020} but simplified to take into account  only a single population. The novelty compared to the standard SEIR model, is that this model takes into account the existence of undocumented/asymptomatic infections, which transmit the virus at a potentially reduced rate. The model tracks the evolution of four  state variables at each day $t$, representing the number of susceptible, exposed, infected-reported and infected-unreported individuals, $S(t), E(t), I^r(t), I^u(t)$ respectively. The parameters of the model are the transmission rate $\beta$ (days$^{-1}$), the relative transmission rate $\mu$ representing the reduction in transmission for asymptomatic individuals, the average latency/incubation period $Z$ (days), the average infectious period $D$ (days) and the reporting rate $\alpha$ representing the proportion of infected individuals which are reported. For a graphic description of the model see Figure~\ref{fig:seir_li_et_al}. The time evolution of the system is defined by the following set of differential equations (recall $N$ denotes the population size):
\begin{equation}
\begin{split}
\frac{d S(t)}{dt}&=-\frac{\beta S(t) I^r(t)}N-\frac{\mu \beta S(t) I^u(t)}N, \\
\frac{dE(t)}{dt}&=\frac{\beta S(t) I^r(t)}N+\frac{\mu\beta S(t) I^u(t)}N - \frac{E(t)}{Z}, \\
\frac{dI^r(t)}{dt}&=\alpha\frac{E(t)}Z-\frac{I^r(t)}D,\\
\frac{dI^u(t)}{dt}&=(1-\alpha)\frac{E(t)}Z-\frac{I^u(t)}D.
\end{split}
\label{eq:Compmodel1}
\end{equation}
Following \cite{Li2020}, we use a stochastic version of this model with a delay mechanism. Each term, say $U$, on the right hand side of \eqref{eq:Compmodel1} is replaced by a Poisson random variable with mean $U$. At each day, we use the 4th order Runge-Kutta numerical scheme to integrate the resulting equations and obtain the values of the four state variables on the next day. For each new reported infection, we draw a Gamma random variable with mean $\tau_d$ days, to determine when this infection will be recorded. For the main analysis we use $\tau_d$=6 days, as the average reporting delay
between the onset of symptoms and the recording of an infection; see
also \cite{Li2020}. Note that the results are robust with respect to the value of reporting delay. The final output of this model is the number of recorded infections on each day $t$, $y=y(t)$. 

\subsubsection{Compartmental Model 2}
\label{sec:compmodel2}
We also use the meta population model of \cite{Li2020}. It models the transmission dynamics in a set of populations, indexed by $i$, connected through human mobility patterns, say $M_{ij}$. This is implemented by incorporating  information on human movement between the 5 main districts of Cyprus: Nicosia, Limassol, Larnaca, Paphos and Ammochostos. In this case, $i=1,2,3,4,5$ and $M_{ij}$ denotes
the daily number of people traveling from district  $i$ to district $j$,
$i \neq j$. Such information is based on the 2011 census data obtained from the Cyprus Statistical Service.
The time evolution of  the four compartmental states in each district $i$ is defined by the following set of differential equations:
\begin{equation}
\begin{split}
\frac{d S_i(t)}{dt}&=-\frac{\beta S_i(t) I_i^r(t)}{N_i}-\frac{\mu \beta S_i(t) I_i^u(t)}{N_i} +\theta\sum_j{\frac{M_{ij}S_j(t)}{N_j-I_j^r(t)}} - \theta\sum_j{\frac{M_{ji}S_i(t)}{N_i-I_i^r(t)}}, \\
\frac{dE_i(t)}{dt}&=\frac{\beta S_i(t) I_i^r(t)}{N_i}+\frac{\mu\beta S_i(t) I_i^u(t)}{N_i} - \frac{E_i(t)}{Z} +\theta\sum_j{\frac{M_{ij}E_j(t)}{N_j-I_j^r(t)}} - \theta\sum_j{\frac{M_{ji}E_i(t)}{N_i-I_i^r(t)}}, \\
\frac{dI_i^r(t)}{dt}&=\alpha\frac{E_i(t)}Z-\frac{I_i^r(t)}D, \\
\frac{dI_i^u(t)}{dt}&=(1-\alpha)\frac{E_i(t)}Z-\frac{I_i^u(t)}D+\theta\sum_j{\frac{M_{ij}E_j(t)}{N_j-I_j^u(t)}} - \theta\sum_j{\frac{M_{ji}E_i(t)}{N_i-I_i^u(t)}},
\end{split}
\label{eq:Compmodel2}
\end{equation}
where the notation follows the notation given in  Sec. \ref{sec:compmodel1}.
In addition to the four state variables, this model also updates at each time step the population of each area $i$, say $N_i$, by
$N_i = N_i + \theta\sum_j{M_{ij}} - \theta\sum_j{M_{ji}}$, where the multiplicative factor $\theta$ is assumed to be greater than 1 to reflect under-reporting of human movement. Like Model \eqref{eq:Compmodel1}, Model
\eqref{eq:Compmodel2} is integrated stochastically using a 4th order Runge-Kutta (RK4) scheme. Specifically, for each step of the RK4 scheme, each unique term on the right-hand side  of the four equations is  determined using a random sample from a Poisson distribution. The
equations describing the evolution of the population in each district $i$, are solved deterministically, $i=1,\ldots,5.$ For more, see
the supplement of \cite{Li2020}.

\subsubsection{Compartmental Model 3}
\label{sec:compmodel3}
Further, we  consider  the meta-population model of \cite{bib:Peng2020}. This is a generalisation of the   classical SEIR model, consisting of seven states:  $(S(t), P(t), E(t), I(t), Q(t), R(t), D(t))$. At time $t$, the {\it susceptible} cases $S(t)$ will become with a  rate $\zeta$ {\it insusceptible} $P(t)$ or with  rate $\beta$ {\it exposed} $E(t)$, that is infected but not yet infectious i.e.  in a latent state. Some of the exposed cases will eventually become {\it infected} with a rate $\gamma$. Infected means  they have the capacity of infecting but are not {\it quarantined} $Q(t)$ yet.  The introduction of the new quarantined state, $Q(t)$, in the classical SEIR model, formed by the infected cases with a constant rate $\delta$, allows to consider the effect of preventive measures. Finally, the quarantined cases, are now split  to {\it cured} cases, $R(t)$, with rate $\lambda(t)$ and to {\it closed}, $D(t)$, with mortality rate $\kappa(t)$.  The model's parameters are the transmission rate $\beta$, the protection rate $\zeta$, the average latent time $\gamma^{-1}$ (days), the average quarantine time $\delta^{-1}$ (days) as well as the time dependent cure rate $\lambda(t)$ and  mortality rate $\kappa(t)$.  The relations are characterized by the following  system of difference equations:

\begin{eqnarray}
    \begin{split}
\label{eq:model3}
\frac{d S(t)}{dt}&=-\frac{\beta S(t) I(t)}N-\zeta S(t), &
\frac{dE(t)}{dt}&=\frac{\beta S(t) I(t)}N-\gamma E(t),\\
\frac{dI(t)}{dt}&=\gamma E(t)-\delta I(t), &
\frac{dQ(t)}{dt}&= \delta I(t)-\lambda(t) Q(t)-\kappa(t) Q(t),\\
\frac{dR(t)}{dt}&= \lambda(t) Q(t), &
\frac{dD(t)}{dt}&= \kappa(t) Q(t),\\ \
\frac{dP(t)}{dt}&=\zeta S(t).
\end{split}
\end{eqnarray}
The total population size is assumed to be constant and equal to
$N=S(t)+P(t)+E(t)+I(t)+Q(t)+R(t)+D(t)$.
According to the official reports, the number of quarantined cases , recovered and deaths , due  to COVID-19, are available. However, the recovered and death cases are directly related to the number of quarantine cases, which plays an important role in the analysis, especially since the numbers of exposed ($E$) and infectious ($I$) cases are very hard to determine. The latter two are therefore treated as hidden variables. This implies that we  need to  estimate  the four parameters $\zeta,\beta, \gamma^{-1}, \delta^{-1}$  and both the time dependent cure rate $\lambda(t)$ and mortality rate  $\kappa(t)$. Notice here that while the rest of the parameters are considered fixed during the pandemic, we allow the cure and mortality rate to vary with time. We expect that the former  will increase with time, given that social distancing measures have been put in place, while the latter will decrease. Finally, this is an optimization problem, and the methodology we have followed in order to address it  can be found in   Appendix \ref{Sec:competailsModel3}.

 \subsubsection{Compartmental Model 4}
\label{sec:compmodel4}

The last model we consider  is a modified version of a solution created by~\cite{bettencourt2008real} to estimate real-time effective reproduction
number  $R_{t}$  \footnote{To avoid confusion, the notation $R_t$  denotes the effective reproduction number but $R(t)$ denotes the number of recovered cases in a given population.}    using a Bayesian approach on a simple Susceptible - Infected (SI) compartmental model:
\begin{equation}
    \begin{split}
\label{eq:model4}
\frac{d S(t)}{dt}&=-\frac{\beta S(t) I(t)}N,\\
\frac{dI(t)}{dt}&=\frac{\beta S(t) I(t)}N-\frac{I(t)}{D}.\\
\end{split}
\end{equation}
We use the Bayes rule to update the beliefs about the true value of $R_t$ based on our predictions and  on how many new cases have been reported each day. Having seen $k$ new cases on day $t$, the posterior distribution of $R_t$ is proportional to (denoted by $\propto$) the
prior beliefs of the value of $P(R_t)$ times the likelihood of $R_t$ given that we have recorded $k$ new cases, i.e.,  $P(R_t | k) \propto P(R_t) \times L(R_t | k)$. To make this iterative  every day that passes by, we use last day's posterior $P(R_{t-1} | k_{t-1})$
to be today's prior $P(R_t)$. Therefore in general $P(R_t | k) \propto \prod_{t=0}^{T}{L(R_t | k_t)}$. However, in the above model the posterior is influenced equally by  all previous days. Thus, we propose a modification suggested in~\cite{systrom2020metric} that shortens the memory and incorporates only the last $m$ days of the likelihood function, $P(R_t | k) \propto \prod_{t=m}^{T}{L(R_t | k_t)}$. The likelihood function is modelled with a Poisson distribution.

\subsection{Estimation of the effective Reproduction Number $R_t$}
\label{sssec:mod12}

Recall the compartmental models discussed in Sec. \ref{sec:compmodel1}
and \ref{sec:compmodel2}. Then the effective reproduction number is given by
\begin{equation}
\label{eq:R}
R_t=\alpha \beta D + (1-\alpha)\mu\beta D,
\end{equation} see the supplement of \cite{Li2020}.
We estimate $R_t$ in \eqref{eq:R} during consecutive fortnight periods for which its value is considered to be constant. To achieve this we estimate the parameters of each model, also assumed to be constant for each fortnight, using daily incidence data for Cyprus \footnote{ In order to maintain consistent notation throughout this article, we use the notation $R_t$, even though in this section the effective reproduction number is considered constant for each fortnight period.}. To estimate the parameters we employ Bayesian statistics, that is, we postulate prior distributions on the parameters and incorporate the data and the model (through the likelihood) to obtain the posterior distributions on the parameters. The posterior distributions capture our updated beliefs about the parameters after combining the prior with the observed data; see, for example, \cite{Bernardo94}.

For the model defined by \eqref{eq:Compmodel1},
we consider the whole area of  Cyprus as a single uniform population. For this case,  the observations are not sufficiently informative to identify all five  parameters of the model. A solution would
be  to enforce identifiability by postulating strongly informative prior distributions on the parameters. Instead, we choose to  make the assumption that the parameters $Z, D$ and $\mu$ have globally constant values, { fixed over time}. In particular we set  $D=3.5$ and $\mu=0.5$ as estimated in \cite{Li2020} and $Z=5.1$ which appears to be the globally accepted mean incubation period. We thus only need to infer the reporting rate $\alpha$ and the transmission rate $\beta$, { which vary both between different fortnights and for} different countries because of the amount of testing and the degree of adherence to the social distancing policies. On the other hand, the model defined by \eqref{eq:Compmodel2} is sufficiently informative to infer all six model parameters.
All computational methods, prior modelling and assumptions in relation to both compartmental models discussed in Sec. \ref{sec:compmodel1} and \ref{sec:compmodel2}
are given in  Appendix \ref{Sec: CompdetailsforR}.
In addition to the above methods we  further consider the method of \citet{Corietal(2013)} as a benchmark to
compare all methodologies for estimating the effective reproduction number.

\section{Results}

\subsection{Descriptive surveillance statistics}\label{ssec:surveillance}

\begin{figure}
\includegraphics[width=0.9\textwidth,trim=0.5cm 7cm 0.5cm 2cm, height=\textheight]{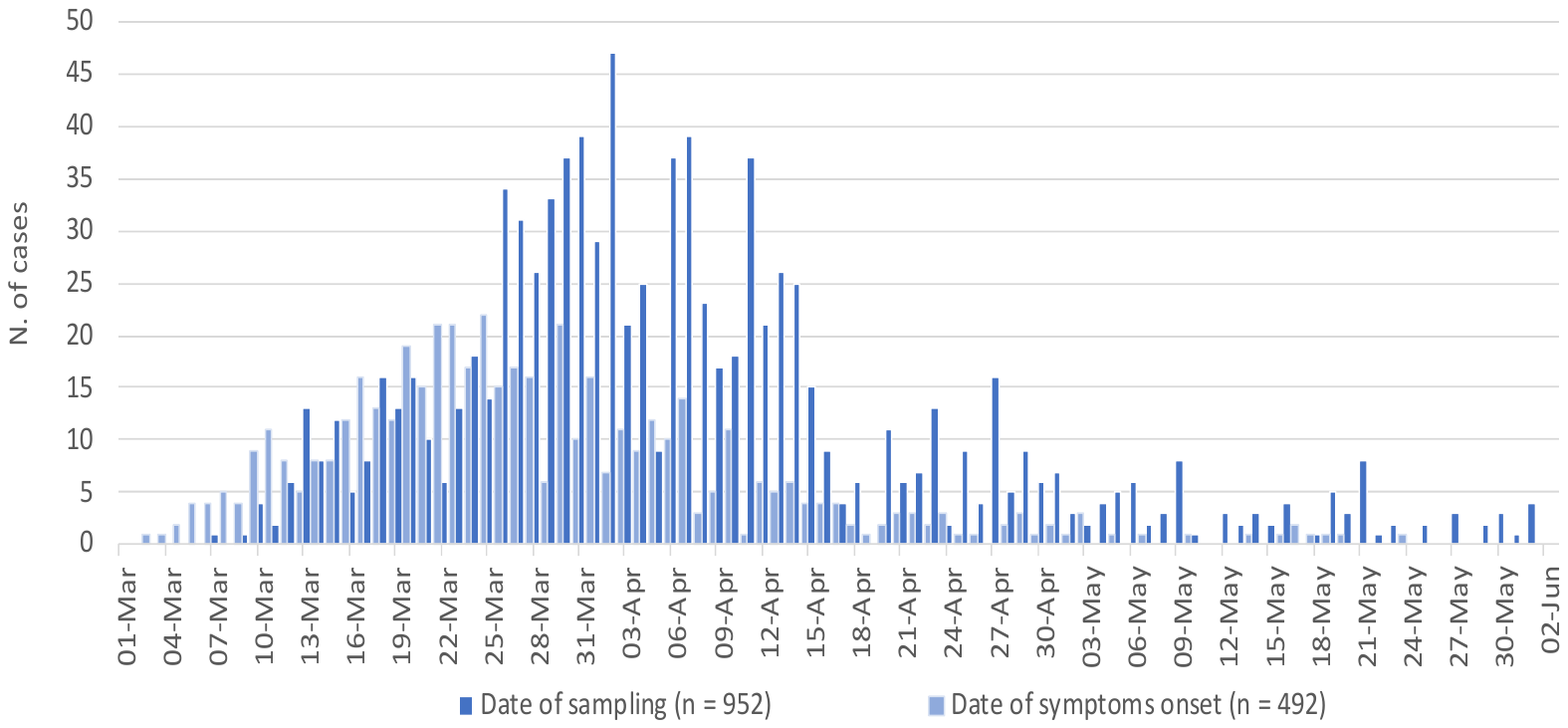}
\vspace{-14cm}
\caption{The epidemic curve by date of sampling and date of symptom onset during the observation period.}
\label{fig:EpidCurve}
\end{figure}

By the end of May 2020, 952 cases of COVID-19 were diagnosed in the Republic of Cyprus. Of these, 50.2\% were male ($n = 478$) and the median age was 45 years (IQR: 31-59 years). The setting of potential exposure was available for 807 cases (84.8\%). Of these, 17.4\% ($n = 140$) had history of travel or residence abroad during a 14-day period before the onset of symptoms. Locally acquired infections were 667 (82.7\%) with 8.6\% ($n = 57$) related to a health-care facility in one geographical setting (cluster A) and 12.4\% ($n = 83$) clustered in another setting (cluster B). The epidemic curve by date of sampling and date of symptom onset is shown in Figure \ref{fig:EpidCurve}. The number of cases started to decline in April reaching very low levels in late May.

\subsection{Long-Term Impact of the COVID-19 epidemic on  Cyprus}
\label{Sec:longtremforecast}

In this section, we investigate the long-term impact of COVID-19 to Cyprus. Towards this, we give long-term projections for the daily incidence and
death rates. We fit system \eqref{eq:model3} to  COVID-19 data that were collected during the period   from the 1st of March 2020 till the 31st of May 2020, in Cyprus. We treat all the reported cases without  making the distinction between {\it local} and {\it imported}. The   model parameters are estimated  using the methodology described in Appendix \ref{Sec:competailsModel3}. Once the model is fitted to data,  it can be used  to forecast the epidemic. In order to study the evolution of the model as  new data are added and the  quality of  the respective forecasts, we have fitted  model \eqref{eq:model3} in four  datasets constructed from the original one using different time periods. Specifically, the four datasets were formed using the    daily  reported incidences from the beginning of the observation period   until and including  the $2/4/2020$, $17/4/2020$,  $15/5/2020$ and  $24/5/202$ respectively. The dates were chosen according  to the change points detected using the methodology described in Section~ \ref{subsub:cpt_detection_methods}, see also Section~\ref{subsub:cpt_detection_application}.

\begin{figure}[H]
\vspace{-4cm}
\centering\includegraphics[width=0.45\textwidth,height=0.60\textheight]{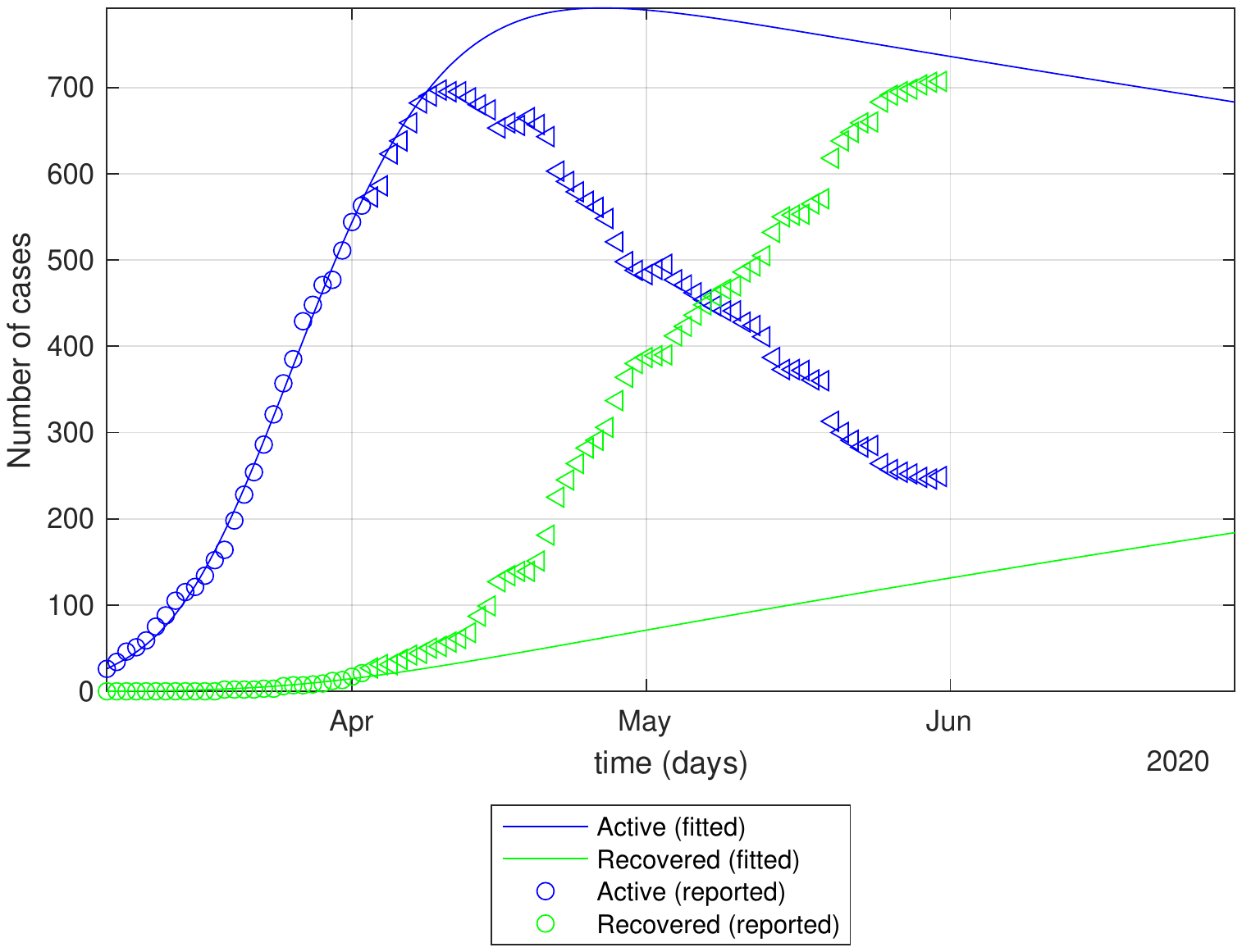}
\centering\includegraphics[width=0.45\textwidth,height=0.60\textheight]{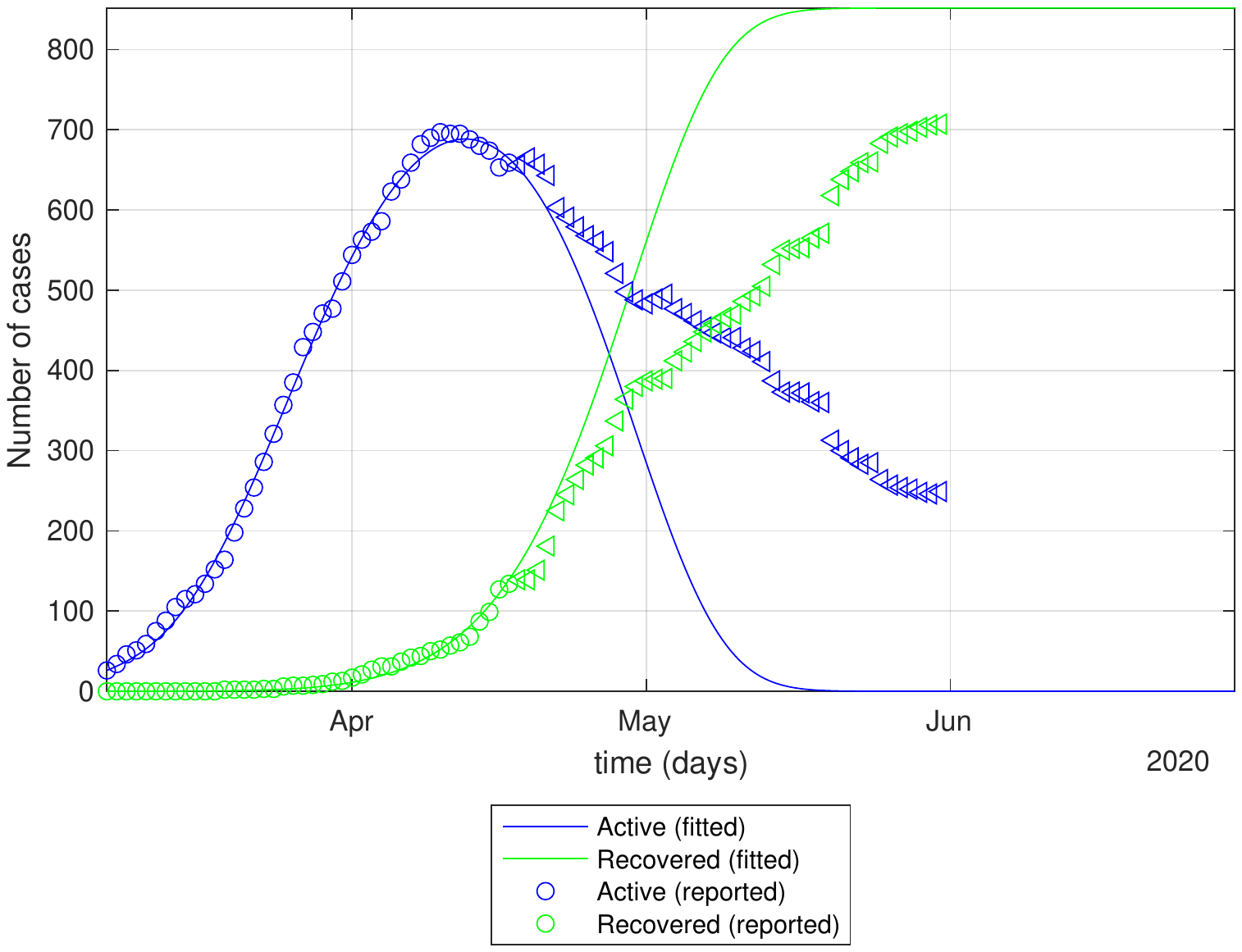}
\\
\vspace{-6cm}
\centering\includegraphics[width=0.45\textwidth,height=0.60\textheight]{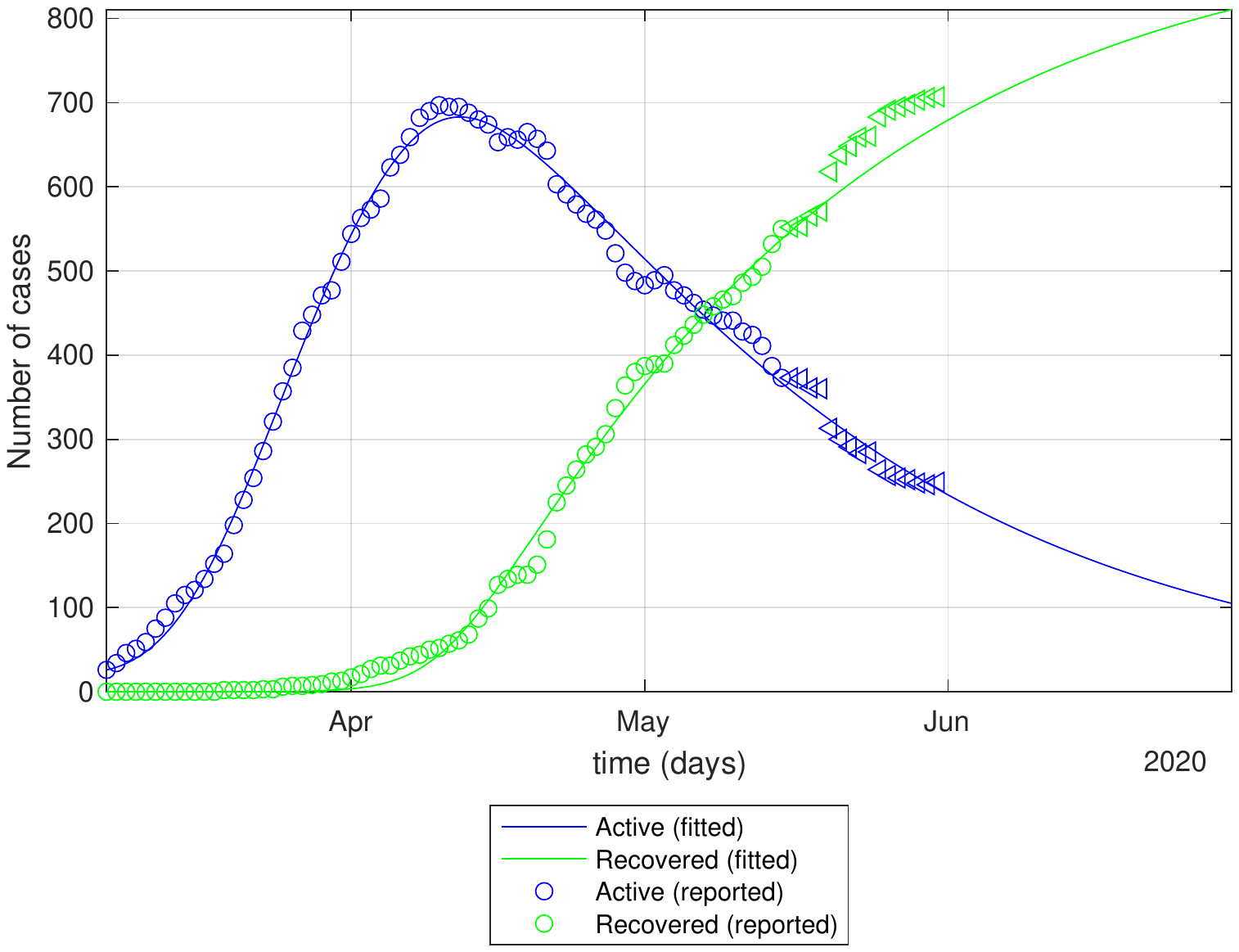}
\centering\includegraphics[width=0.45\textwidth,height=0.60\textheight]{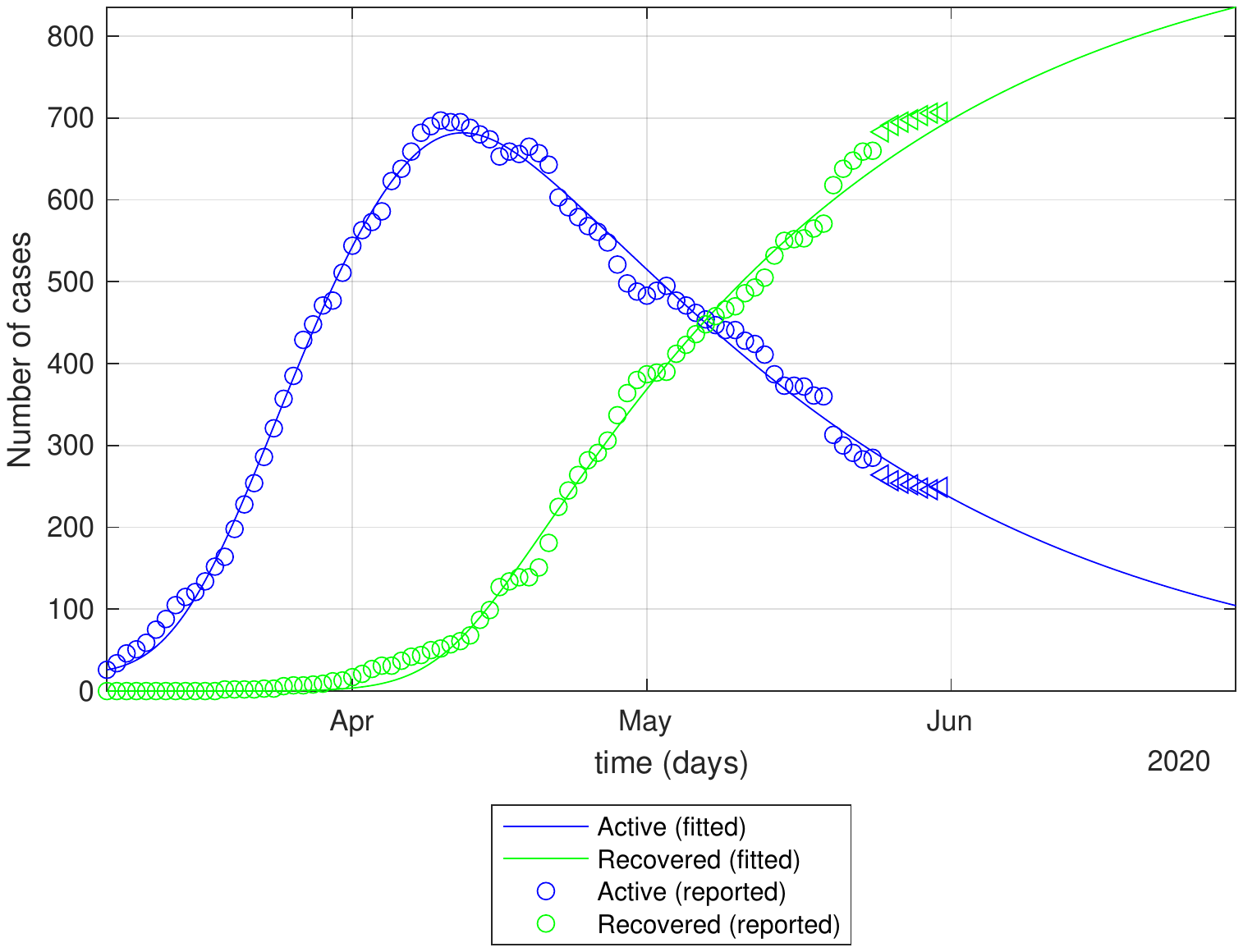}
\vspace{-4cm}
\caption{Predictions obtained after fitting  \eqref{eq:model3} for  the  number of active  cases (green) and recovered cases (blue) until the end of the observation period. Parameter estimation was performed  by subsampling the data until  (top left)  2/4/2020, (top right), 15/4/2020, (bottom left) 15/05/2020  (bottom right) 24/5/2020.The relative error   equals  $0.4\%, 0.2\%, 0.3\%, 0.3\%$ for the recovered and  $18\%, 5.8\%, 0.16\%, 0.1\%$ for the active cases.}
\label{fig:pred_model3}
\end{figure}

The fitted model in each case was used in order to predict the pandemic's evolution until the $30/6/2020$. In
Figure \ref{fig:pred_model3}, we show the number of  predicted exposed  plus infectious cases (green solid lines) and the number of predicted recovered cases (blue solid lines)   for the duration  of the prediction period, and compare them to the observed cases which are indicated by  circles and triangles. We use circles for  data that have been used in the prediction and  triangles for the observed data that are used for validation. Visual inspection shows that after a period of about two months during which  the  model overestimates the number of active cases and underestimates the number of recovered, see Figure \ref{fig:pred_model3} (top), model \eqref{eq:model3} was able to capture accurately the evolution of the pandemic,  Figure \ref{fig:pred_model3} (bottom).

The performance of the predictions can also be evaluated  by means of  the relative error (RE) which are computed using $RE=\sqrt{\frac{\sum_{t}(y_t-x_t)^2}{\sum_{t} x_t^2 }}$, where $x_t$
denotes the datum  for day $t$ and $y_t$ the model prediction for the same day. The RE for the recovered cases equal $0.4\%, 0.2\%, 0.3\%$ and $0.3\%$ for the four time periods respectively with the corresponding RE for the active cases being  high in the beginning  $18\%, 5.8\%$, but then dropping considerably  $0.16\%$ and $0.1\%$, reflecting the fact the model caught up with the evolution of the pandemic. Overall, system \eqref{eq:model3} gives adequate predictions especially when data from longer time
periods are used.

\begin{figure}[H]
\vspace{-4cm}
\centering\includegraphics[width=0.45\textwidth,height=0.65\textheight]{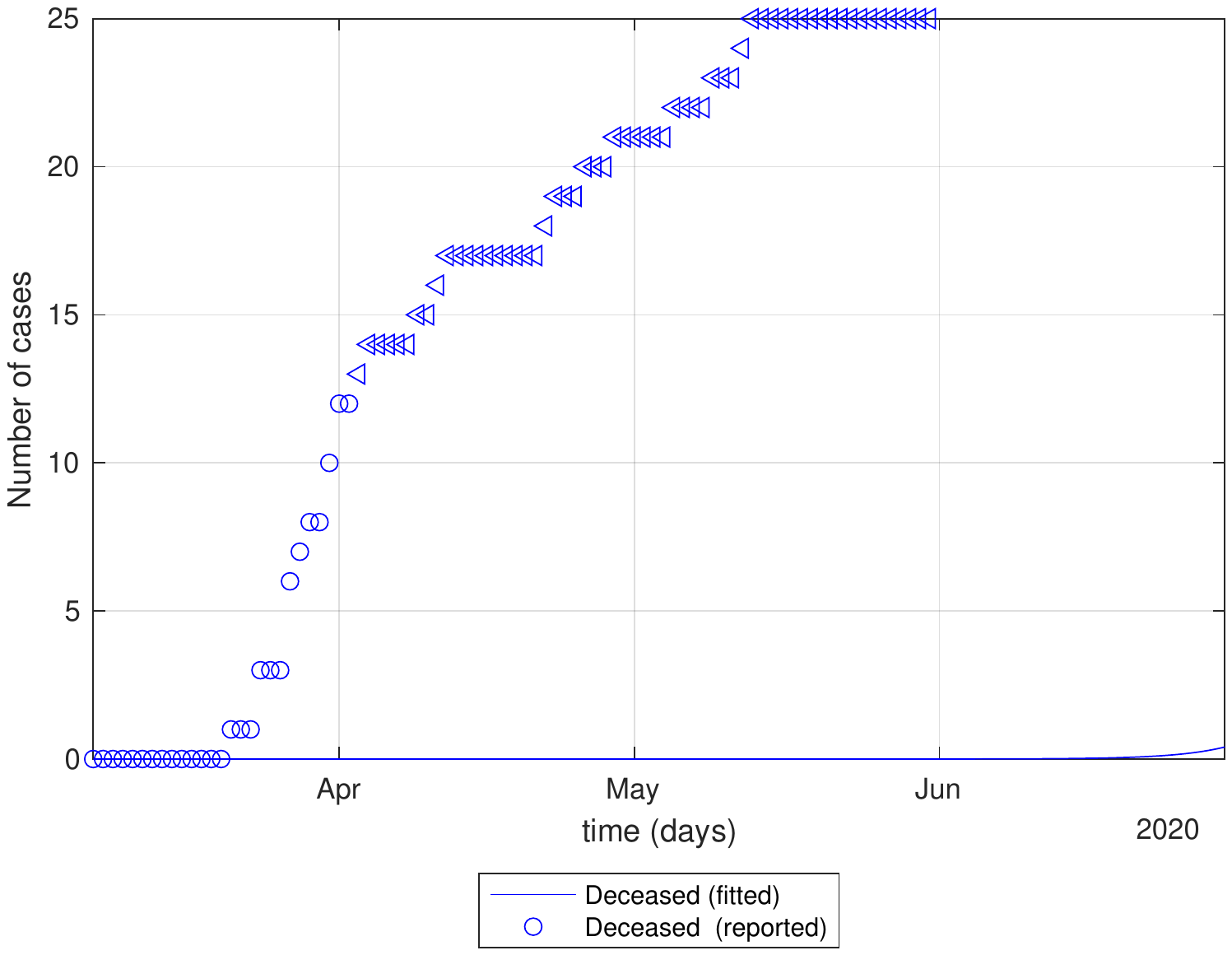}
\centering\includegraphics[width=0.45\textwidth,height=0.65\textheight]{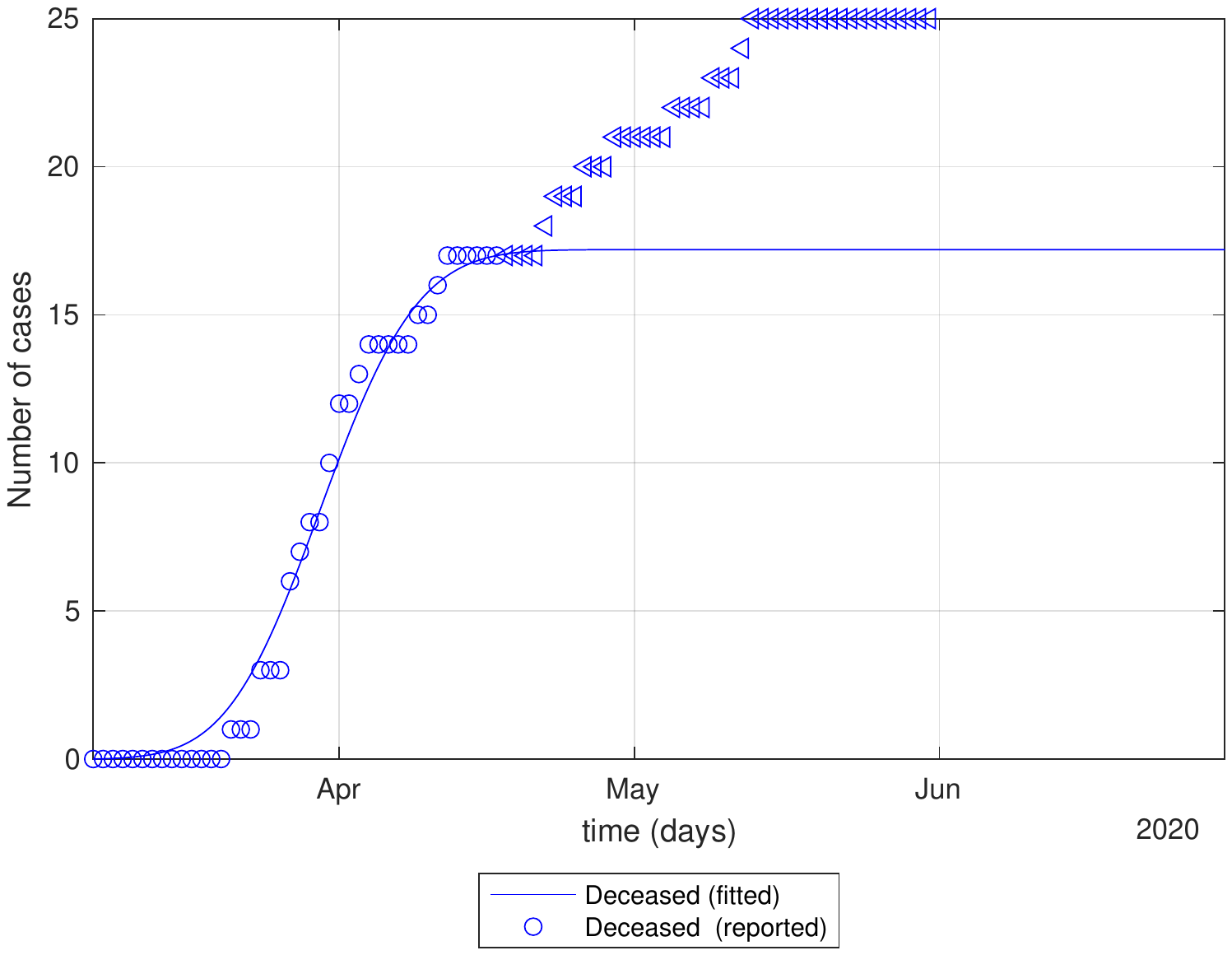}
\\
\vspace{-6cm}
\centering\includegraphics[width=0.45\textwidth,height=0.65\textheight]{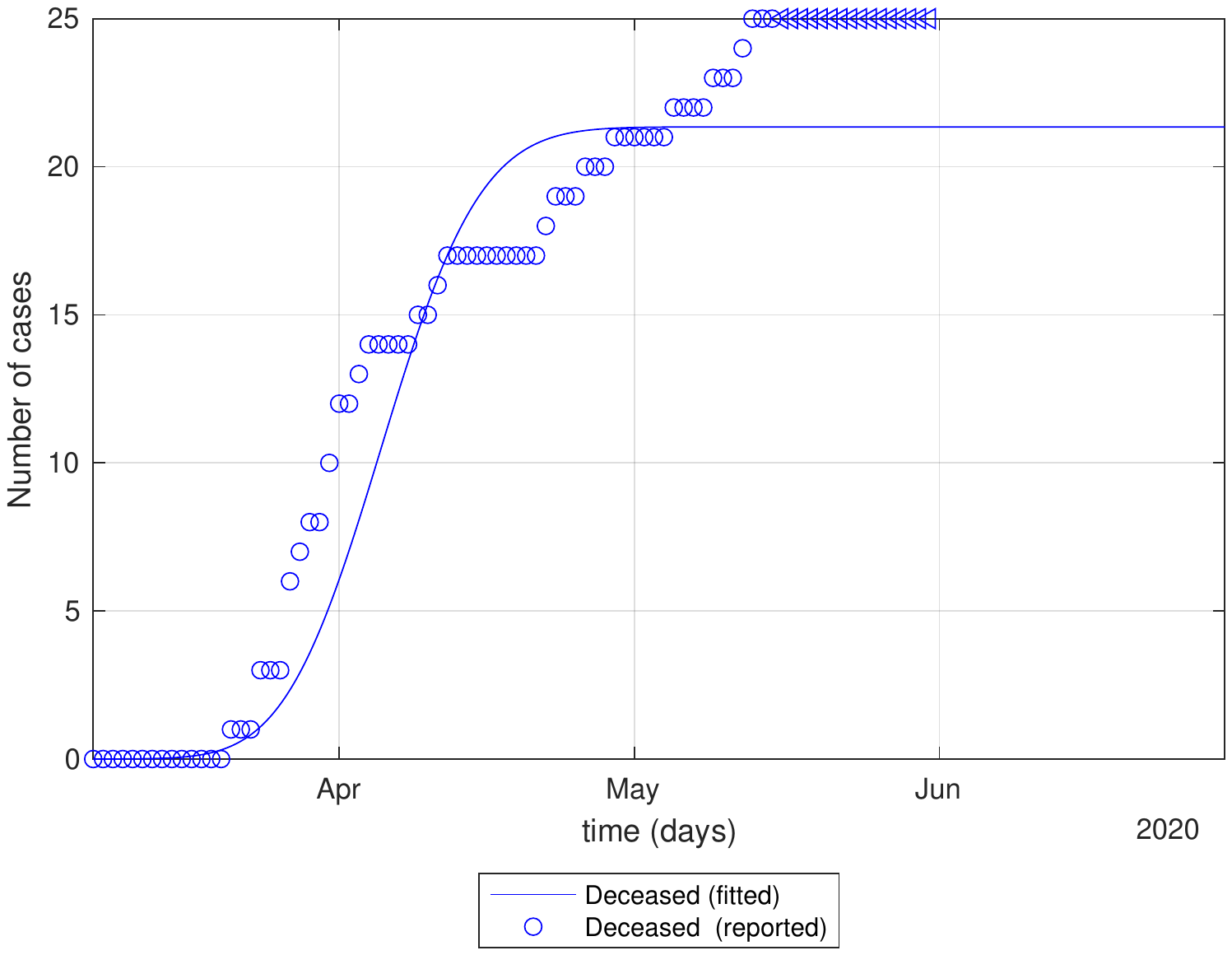}
\centering\includegraphics[width=0.45\textwidth,height=0.65\textheight]{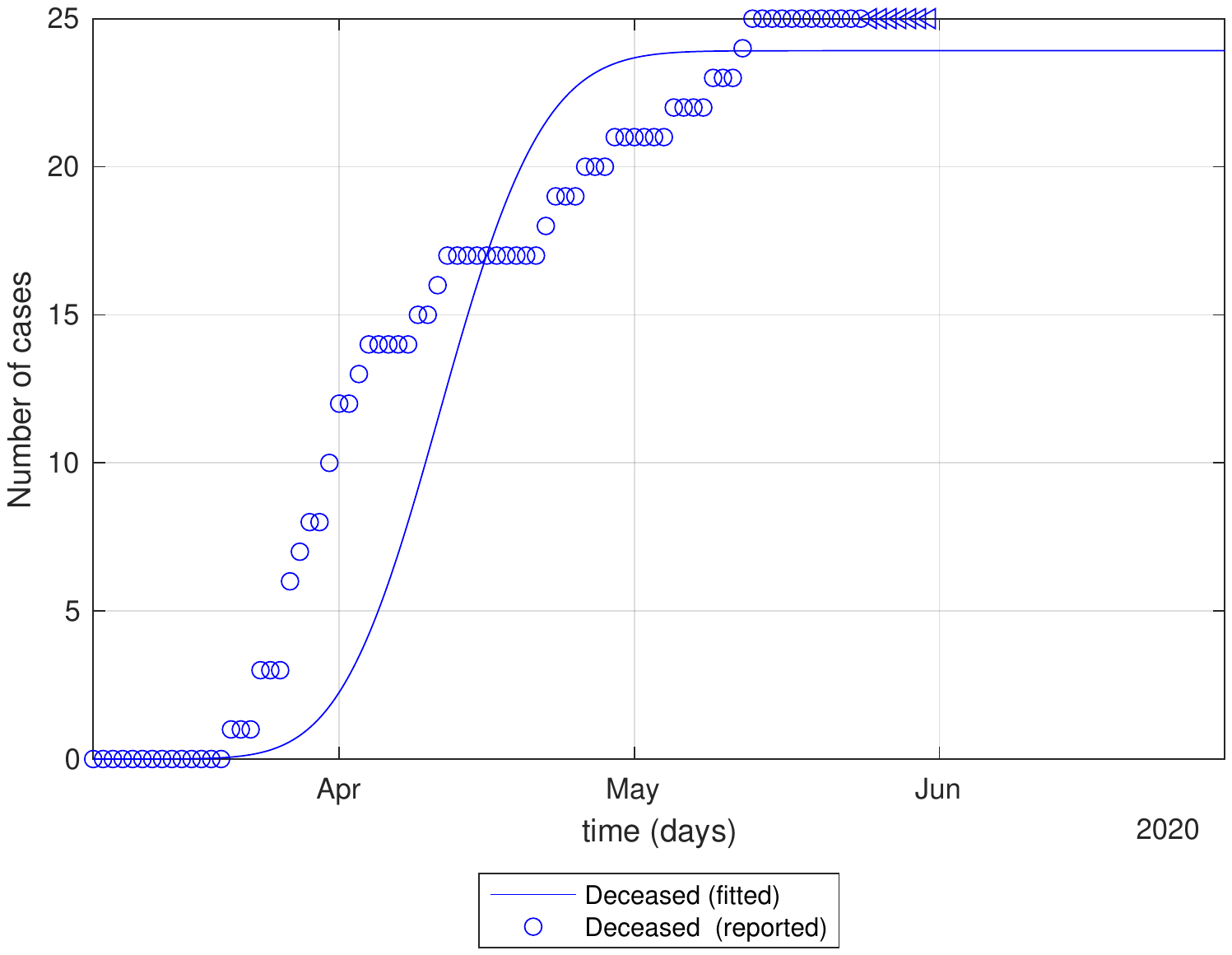}\vspace{-4cm}

\caption{Predictions obtained after fitting  \eqref{eq:model3} for  the  number of deaths  until the end of the observation period. Parameter estimation was performed  by subsampling the data until    2/4/2020 (top left),  15/4/2020 (top right),  15/05/2020 (bottom left),   24/5/2020 (bottom right). Corresponding RE's equal $2\%, 4\%$ and $8\%$  clockwise from top right to bottom right. }
\label{fig:death_model3}
\end{figure}

Figure \ref{fig:death_model3} shows the number of deaths and their
respective predictions using subsets of  data as described above.  In the duration of the first data set, there were no deaths registered and therefore the prediction was identically zero, giving also an RE equal to $100\%$ see Figure \ref{fig:death_model3} (top left). As more deaths are registered the model's ability to predict the correct number of deaths is improving, see  Figure  \ref{fig:death_model3}. 

\begin{figure}[H]
\vspace{-3cm}
\centering\includegraphics[width=0.45\textwidth,height=0.65\textheight]{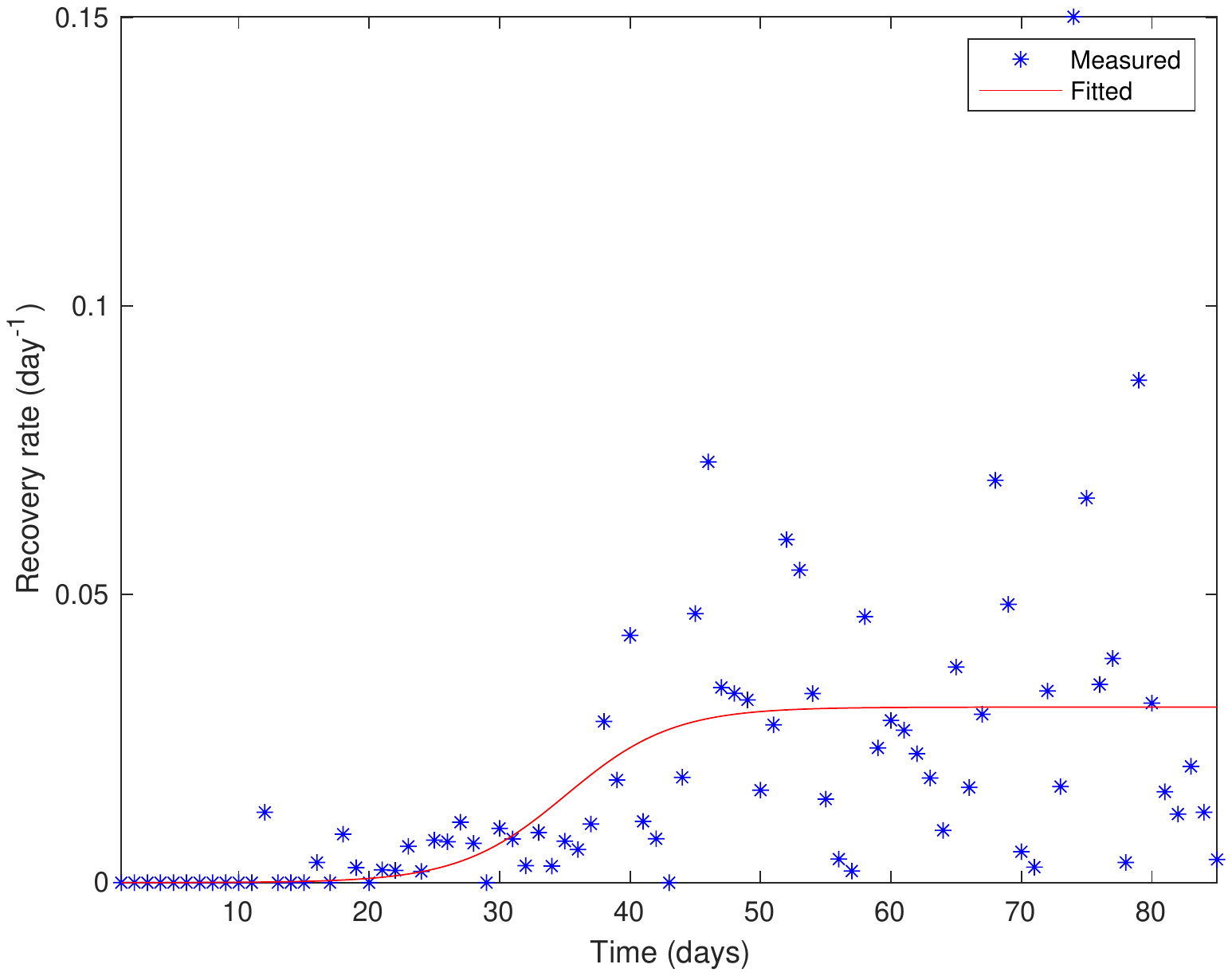}
\centering\includegraphics[width=0.45\textwidth,height=0.65\textheight]{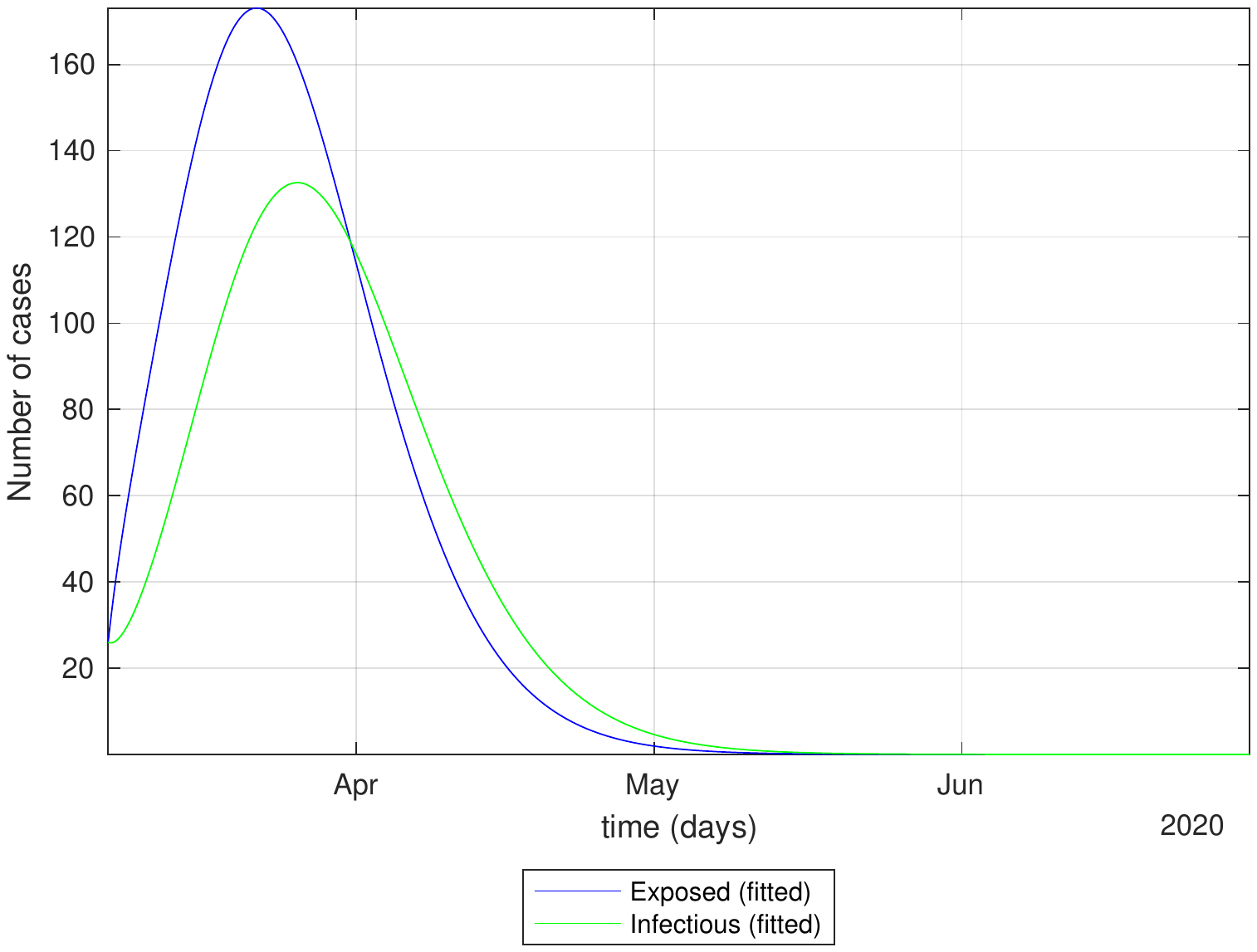}
\vspace{-2cm}
\caption{ (Left plot ) The recovery rate (days$^{-1}$) (blue stars) and the corresponding fitted function \eqref{eq:recovery}  (red solid line).
(Right plot) Exposed $E(t)$ and infectious $I(t)$ cases.}
\label{fig:recovery_model3}
\end{figure}

The recovery rate ($\lambda(t)$) is modelled as
\begin{equation}
\label{eq:recovery}
\lambda(t)=\frac{\lambda_1}{1+\exp(-\lambda_2(t-\lambda_3))}, \quad \lambda_i\ge 0, \quad i=1,2,3.
\end{equation}
The idea is that the recovery rate, as time increases, should converge towards a constant. In Figure \ref{fig:recovery_model3} (left), the fitted recovery rate (solid line) is  plotted against the observed number of recovered cases  (stars).

Finally,  Model 3 can be used to  estimate the unobserved number of exposed, $E(t)$, and infectious, $I(t)$, cases during the development of the pandemic. The maximum number of exposed cases occurs on the 21st of March 2020 and is estimated to be 173 cases, Figure \ref{fig:recovery_model3} (right, blue line), with the maximum of infectious individuals (136) being attained on the 26 of March 2020. We can observe a delay in the transition of exposed to infectious in the order of 5 days, which  suggests a 5 day latent time of COVID-19.

\subsection{Change-point Analysis}
\label{subsub:cpt_detection_application}

\begin{figure}[htb]
\centering\includegraphics[width=0.8\textwidth,height=0.33\textheight]{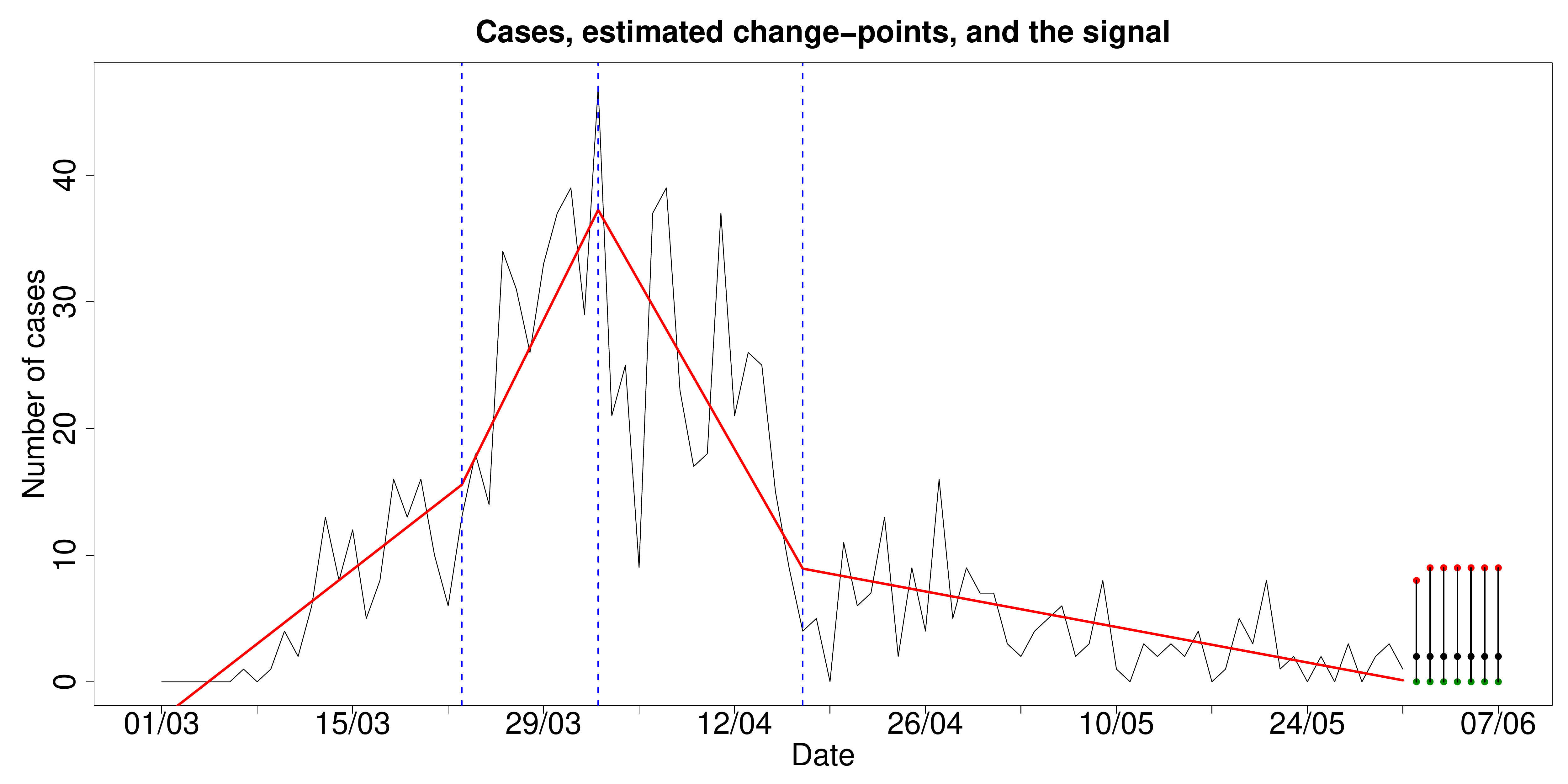}
\caption{The real data (black coloured line) and the estimated piecewise-linear fit (red coloured line) for the daily incidence rate. The change-point locations are given with dotted, blue vertical lines. At the right-end of the plot  point estimators (black dots) and $95\%$ prediction intervals for the number of daily cases for the next week  are reported.}
\label{sampling_cases_linear}
\end{figure}

We first consider the change-point detection method of Sec.
\ref{subsub:cpt_detection_methods} for the case of piecewise-linear
signal plus noise model.  Figure \ref{sampling_cases_linear}
illustrates the results obtained by this analysis on daily incidence data.

This  method has detected three important changes. The first change occurs on the $23^{rd}$ of March, while the last two changes are estimated on the $2^{nd}$ and $17^{th}$ of April. The first change-point, on the $23^{rd}$ of March, indicates  a significant increase in the number of cases, and it is believed to be related to the development of the clusters A and B as mentioned in Section \ref{ssec:surveillance}. The second change-point shows that the upward trend vanishes and a negative slope takes its place leading to a vast decrease in the number of cases. There is a connection of this change point with the Government’s lockdown  decrees on the $24^{th}$ and $31^{st}$ of March; it shows the almost immediate impact that such decisions can have in fighting the pandemic. The third change-point indicates
stabilisation in the number of the detected cases over  the last period. The median for the number of cases in the last segment is equal to 3.
Predictions (together with 95\% C.I.) for the week ahead are also shown in Figure  \ref{sampling_cases_linear}. Note that there will be no more than 8 cases on the $1^{st}$ of June and no more than 9 cases per day in the period from the $2^{nd}$ of the month until the $7^{th}$. The point estimator is found to be equal to 2 cases per day.  The corresponding analysis for the  piecewise-constant signal plus noise
model is given in Appendix \ref{Sec: Piecewise Constant}. Both models
describe adequately the daily time series of number of cases and
(a) they provide a justification of the positive impact of the Government's measures in fighting COVID-19, (b)  they give a clear view of how contagious the virus is, especially in cases when hubs are developed in the society, and (c) the  last homogeneous discovered segment provides  better understanding of the current state of the virus in terms of its transmittal.

\subsection{Count Time Series Analysis}

\begin{figure}[!ht]
\centering
\includegraphics[width=\textwidth,height=0.33\textheight]{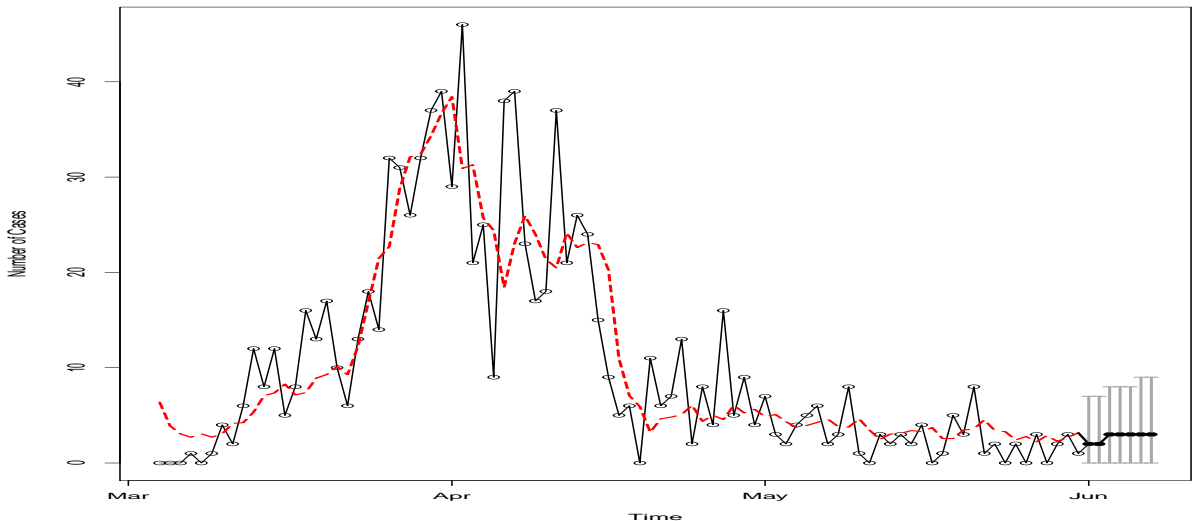}
\caption{Daily number of COVID-19 cases  in Cyprus
from 01/03/2020  to 31/05/2020 (black line) and the fitted model with
interventions (red line). The grey bars at the end of the plot show 95\% simultaneous prediction intervals.}
\label{fig:tscount}
\end{figure}

We first fit model  \eqref{eq:log-linear model} without interventions to daily incidence data, considering the time  period between the 4th of March and  the 31st of May. Maximum likelihood estimation shows that the fitted model is given by
\begin{eqnarray*}
\hat{\nu}_{t} = -0.003_{(.009)} + 0.547_{(0.447)} \hat{\nu}_{t-1}+
0.451_{(.050)} \log(1+Y_{t-1}).
\end{eqnarray*}
Corresponding standard errors of the regression coefficients are  given underneath in parentheses. Note that the sum of the coefficients $0.547+0.451 \sim 1$ which
shows evidence of some non-stationarity observed in the data. Exploring further the data by investigating existence of interventions we find two additive outliers on 13th and 26th of March ($p$-value after adjusting for all type
of interventions is negligible). The resulting model is given by
\begin{eqnarray*}
\hat{\nu}_{t} =-0.007_{(.007)} + 0.779_{(0.046)} \hat{\nu}_{t-1}
+0.211_{(.046)}  \log(1+Y_{t-1})+1.643_{(0.137)} I(t=10) + 1.102_{(0.288)}
I(t=23).
\end{eqnarray*}
Note again that the sum $0.779+211 \sim 1$ which shows that the non-stationarity
persists even after including additive outliers (in the log-scale). Furthermore,
the positive sign of both interventions shows the sudden explosion of the daily number of people infected.  The corresponding BIC values obtained after
fitting this model is equal to 576.643 which improves the BIC of the
model without intervention which was equal to 615.766.
Figure \ref{fig:tscount} shows the fit of the model to the data and gives 95\% prediction intervals for the week ahead.
Comparing both change-point analysis (see Fig.
\ref{sampling_cases_linear}) and the result obtained by using the above
intervention analysis, we observe that both approaches give similar prediction  intervals that include future observed incidence data.  Indeed, the observed data for the week ahead (01/06/2020-07/06/2020) were 4,6,1,0,5,5 and 1 cases.

\subsection{Results for {\bf the effective reproduction number}}

Recall the effective reproduction number $R_t$ defined by \eqref{eq:R}.
We perform Bayesian analysis using \eqref{eq:Compmodel1} (see Section \ref{sssec:mod12}) separately on two sets of data: the data concerning all COVID-19 incidents in Cyprus and the data concerning incidents from local transmission only; for details regarding prior modelling and computation see Appendix \ref{Sec: CompdetailsforR}. For each fortnight period under examination, we draw 10000 steps of the used Markov chain Monte Carlo algorithm (independence sampler), discard the first 2000 steps as burn-in and use the remaining ones to approximate the posterior distributions of $\alpha, \beta$ and hence $R_t$.

\begin{figure}[H]
\centering\includegraphics[width=0.8\textwidth]{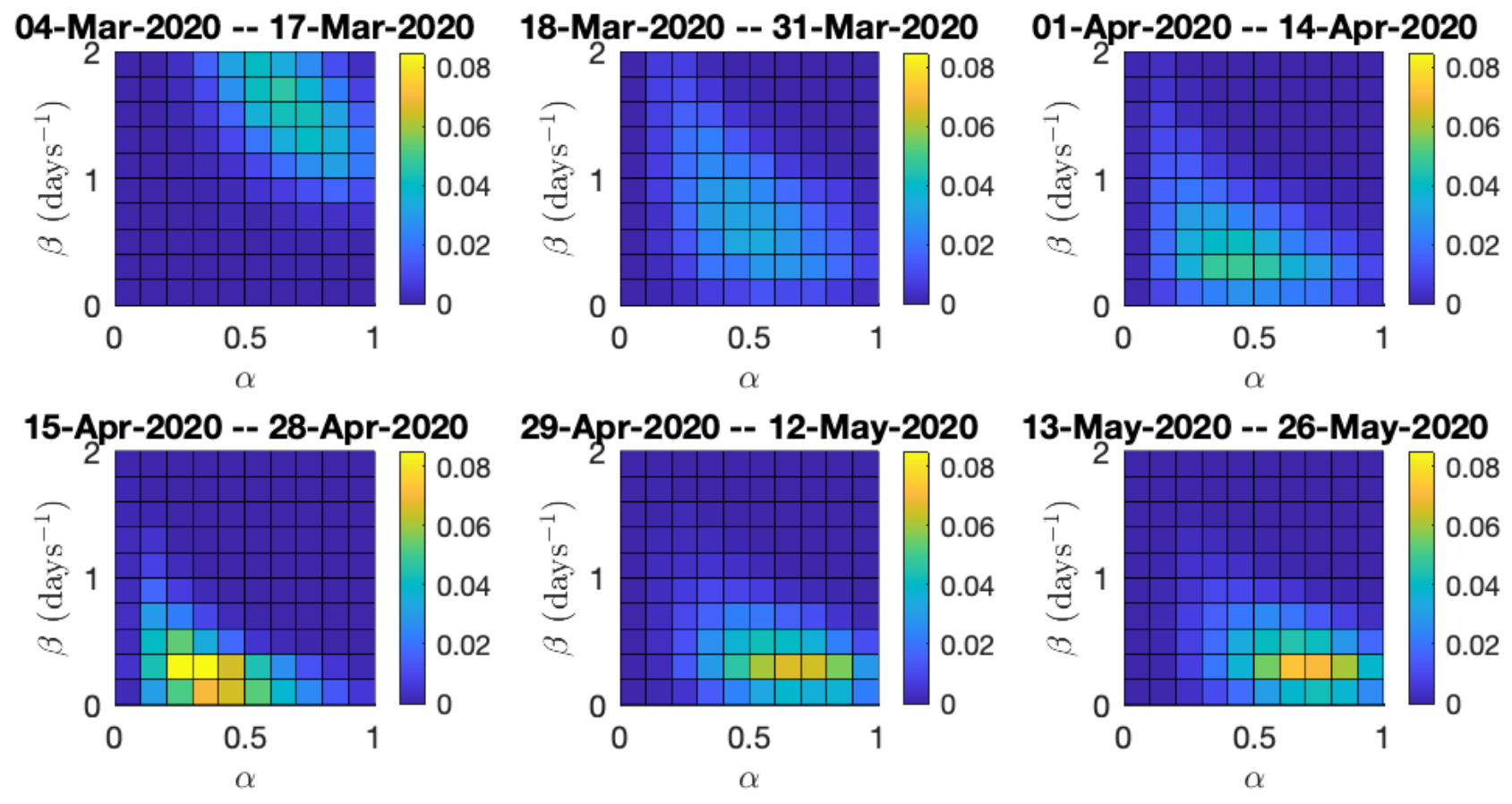}
\caption{Joint posterior distributions of the reporting rate $\alpha$ and the transmission rate $\beta$ in Model 1, for the 6 fortnight periods starting from $04/03/2020$ until $26/05/2020$. Analysis using the full data.}
\label{ab_post}
\end{figure}

For the data concerning all incidents, the first recorded incident was on 07/03/2020, hence, as detailed in Appendix \ref{Sec: CompdetailsforR}, we initialize our analysis of the outbreak  3 days earlier, on 04/03/2020. Figure \ref{ab_post} shows the estimated joint posterior distributions of the reporting rate $\alpha$ and the transmission rate $\beta$, for the six fortnight periods. In the first period the posterior probability is distributed in a wide range of values, while in the following periods it concentrates on a narrower range of values. In particular, in the first period, $\beta$ takes with high posterior probability values between 1 and 2, while $\alpha$ concentrates between 0.5 and 1. This can be attributed on both the data and the postulated priors on $\alpha$ and $\beta$. Early in the outbreak, there was a high degree of social interactions, and only a small fraction of the public took protective measures, hence an infected individual could transmit the virus to many people before getting detected and put into isolation; this explains the high values of $\beta$. Furthermore, there were only a few recorded incidents, the virus had not yet spread in the community and due to the effective contact tracing procedures, the reporting rate was high; this can explain the high values of $\alpha$ (even though the prior on $\alpha$ at this stage penalizes values far from 0.5). In the next two periods, the introduction of lockdown measures by the government and the adherence of the majority of the public to the advised protective measures, results to lower values of $\beta$. At the same time, the virus has penetrated certain local communities (cf. hubs A and B in Subsection \ref{ssec:surveillance}),
 and as a result we have lower values of the reporting rate $\alpha$, initially around $0.5$ and in the third and fourth periods between 0.2 and 0.5 (even though the prior on $\alpha$ at this stage puts higher penalty on small values). In the fourth period, there is a high concentration of the posterior distribution on even lower values of $\beta$. This is the effect of the continued strict lockdown imposed. Finally in the final two periods, the values of $\beta$ remain low, with a slight increase compared to the fourth period, which can be attributed to the relaxation of measures by the government on 04/05/2020. The values of the reporting rate $\alpha$ significantly increase in the last two periods, which can be attributed to the very high number of both targeted and random testing performed.

\begin{figure}[H]
\centering\includegraphics[width=0.7\textwidth]{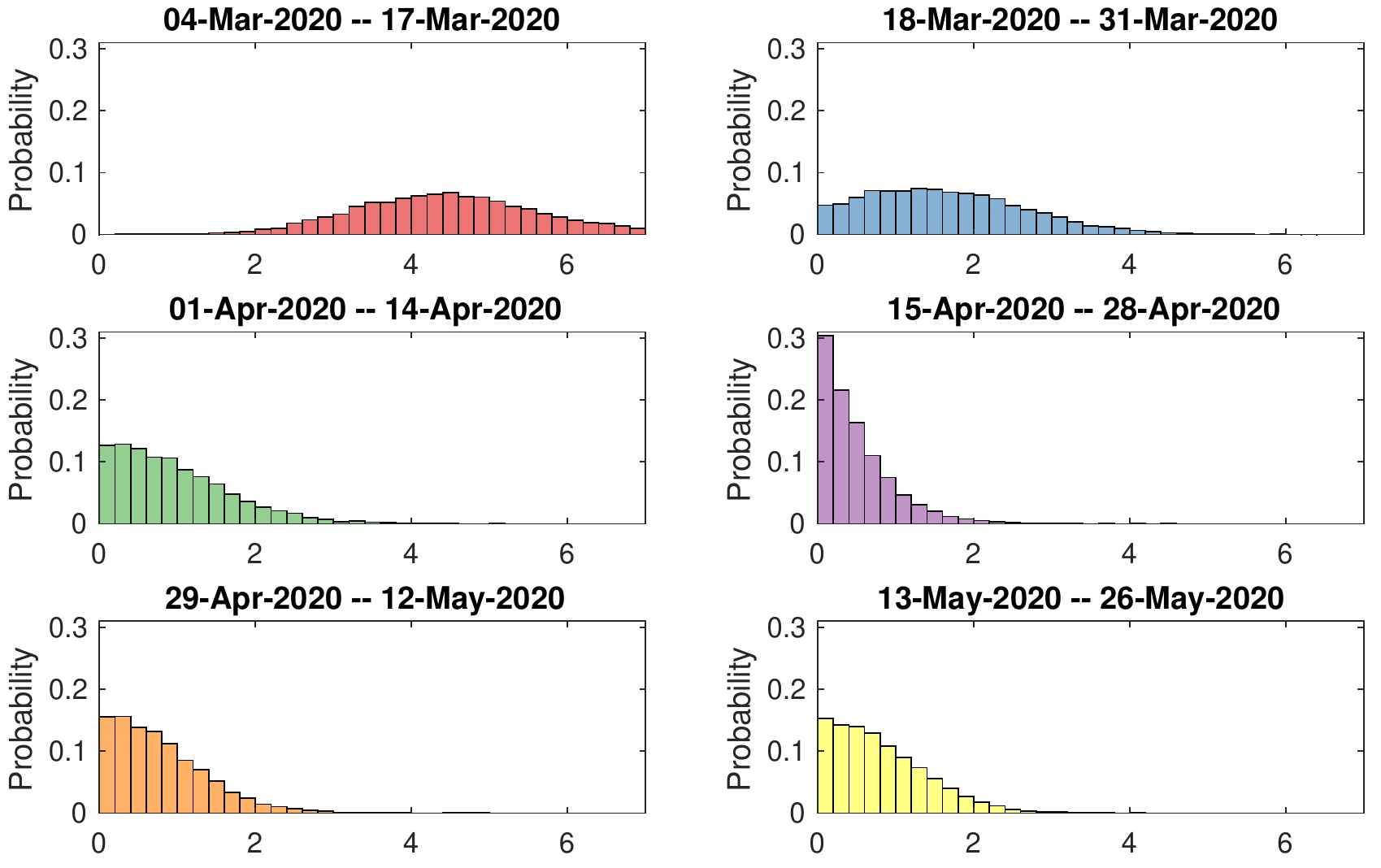}
\caption{Posterior distributions of the effective reproductive number in Model 1, for the six fortnight periods starting from $04/03/2020$ until $26/05/2020$. Analysis using the full data.}
\label{R_hists}
\end{figure}

\begin{figure}[H]
\centering\includegraphics[width=0.7\textwidth]{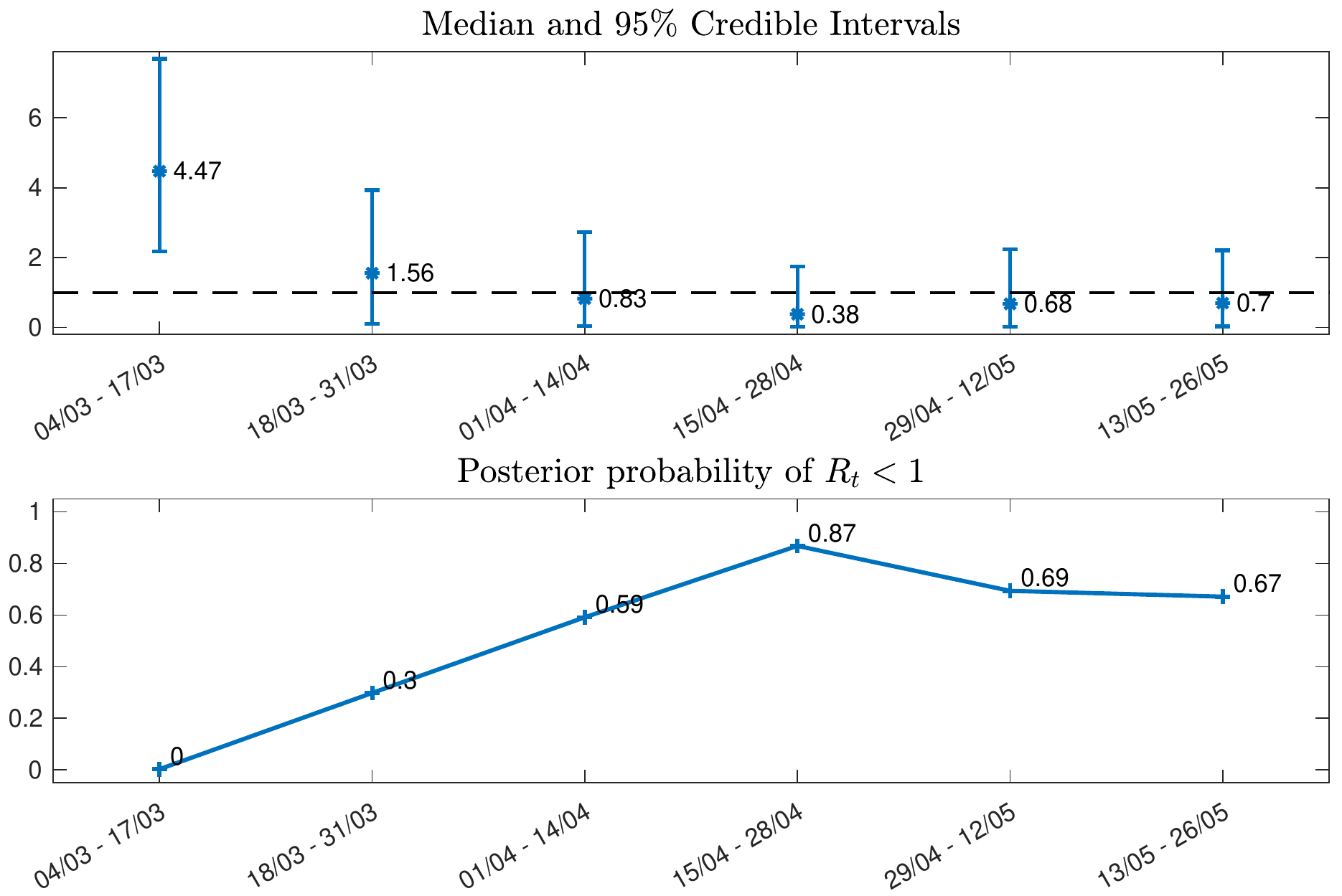}
\caption{Top: median and 95$\%$ credible intervals for the posterior distributions of the effective reproduction number in Model 1, for the six fortnight periods starting from $04/03/2020$ until $26/05/2020$. Bottom: posterior probabilities of the event $R_t<1$. Analysis using the full data.}
\label{R_estimates}
\end{figure}

The considerations in the last paragraph, combined with equation \eqref{eq:R}, explain the results on the effective reproduction number $R_t$ in Figures \ref{R_hists} and \ref{R_estimates}. In Figure \ref{R_hists} and the top part of Figure \ref{R_estimates}, we see that the posterior distribution of $R_t$, is spread in a wide range of high values between 3-6, with a median of 4.47, while in the next three periods with the introduction of the progressively stricter measures the posterior increasingly concentrates on lower values. In particular in the fourth period, the posterior median is 0.38. In the last two periods, with the relaxation of the lockdown, there is a small increase in the values of $R_t$, however its posterior distribution still mostly concentrates under 1, with medians around 0.7. The bottom part of Figure \ref{R_estimates}, shows the posterior probabilities of the event $R_t<1$, which steadily increase following  the progressively stricter measures imposed, from 0 in the first period to 0.87 in the fourth period, before slightly dropping to around 0.69 and 0.67 with the relaxation of measures in the last two periods, respectively.

For the data concerning only locally transmitted incidents, the first recorded incident was on 10/03/2020, hence we initialize our analysis of the outbreak  3 days earlier, on 07/03/2020.
The results of the analysis using data concerning only local transmission incidents, are similar to the ones of the analysis using the full data;
see Figures \ref{ab_post_local}, \ref{R_hists_local} and \ref{R_estimates_local}
in Appendix.




Next, we consider the estimation model described in~\cite{Li2020} where Cyprus is divided in 5 subpopulations (Nicosia, Limassol, Larnaca, Paphos, Ammochostos) and the mobility patterns between them are taken into account (as described in Metapopulation compartmental model 2). The effective reproduction number is given by \eqref{eq:R}. The compartmental model 2 structure was integrated stochastically using a 4th order Runge-Kutta (RK4) scheme. We use uniform prior distributions on the parameters of the model, with ranges  similar to ~\cite{Li2020} as follows: relative trasmissibility $0.2\leq\mu\leq 1$, movement factor $1\leq\theta\leq 1.75$; latency period  $3.5\leq Z\leq 5.5$; infectious period $3\leq D \leq 4$. For the infection rate we choose $0.1\leq\beta\leq1.5$ before the lockdown and $0\leq\beta\leq0.8$ after the lockdown and for the reporting rate we choose $0.3\leq\alpha\leq1$. Note that the Ensemble Adjustment Kalman Filter (EAKF, described in Appendix \ref{Sec: CompdetailsforR}) is not constrained by the initial priors and can migrate outside these ranges to obtain system solutions. For the initialization purposes we assume that all 5 districts are potential origins with an undocumented infected and exposed population drawn from a uniform distribution $[0,5]$ a week before the first documented case. Initial condition does not affect the outcome of the inference. Transmission model 2 does not explicitly represent the process of infection confirmation. Thus, we mapped simulated documented infections to confirmed cases using a separate observational delay model. In this delay model, we account for the time interval between a person transitioning from latent to contagious and observational confirmation of that individual infection through a delay of $T_d$. We assume that $T_d$ follows a Gamma distribution $G(a,\tau_d/a)$ where $\tau_d=6$ days and $a=1.85$
as derived by~\cite{Li2020} using data from China. Inference is robust with respect to the choice of $\tau_d$.

\begin{figure}[H]
\centering\includegraphics[width=0.5\textwidth]{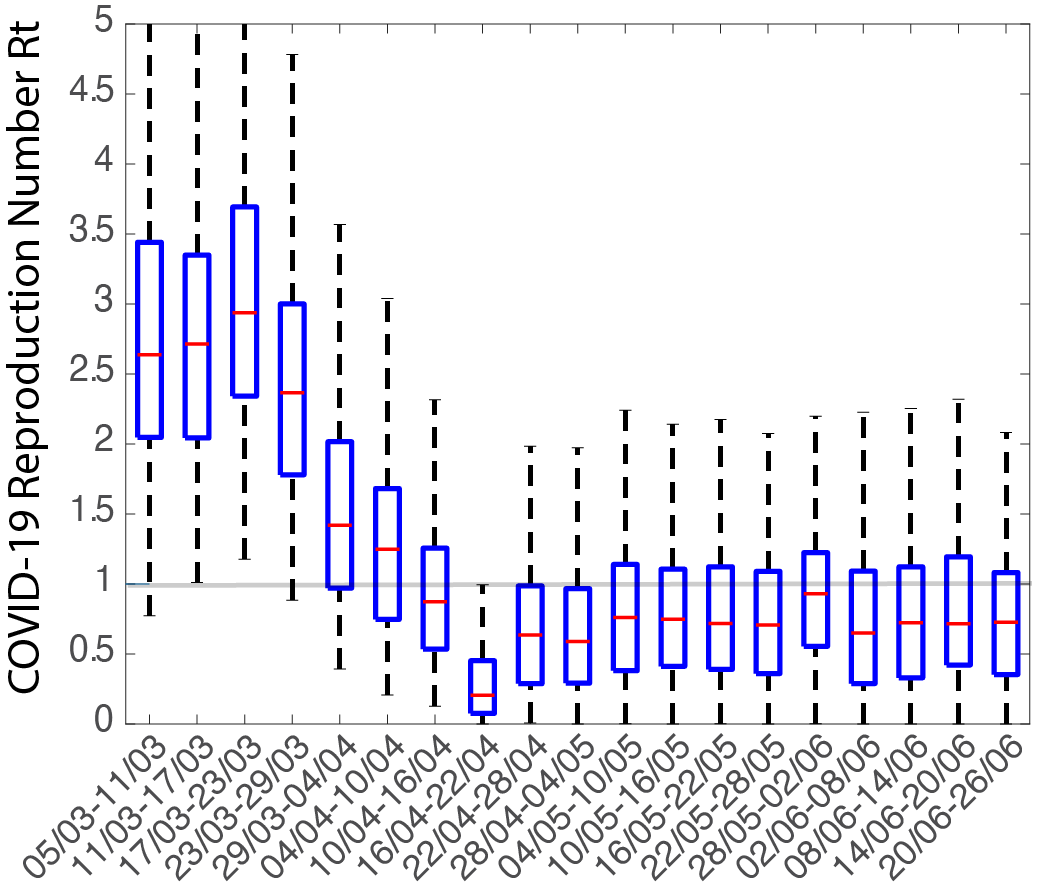}
\caption{Median value (red) and 95$\%$ credible intervals (box) for the posterior distributions of the effective reproduction  number in metapopulation compartmental model 2, for the period between $05/03/2020$ until $26/06/2020$.}
\label{model2_Rt}
\end{figure}

For the inference we use incidents from local transmission in Cyprus as were reported by the Ministry of Health. In Figure \ref{model2_Rt} we plot the time evolution of the weekly effective reproduction number $R_t$. While at the beginning of the outbreak the effective reproduction number was close to 2.5, after the lockdown measures, it  dropped below 1 and stayed consistently there until the end of June 2020.

\begin{figure}[H]
\centering\includegraphics[width=0.70\textwidth]{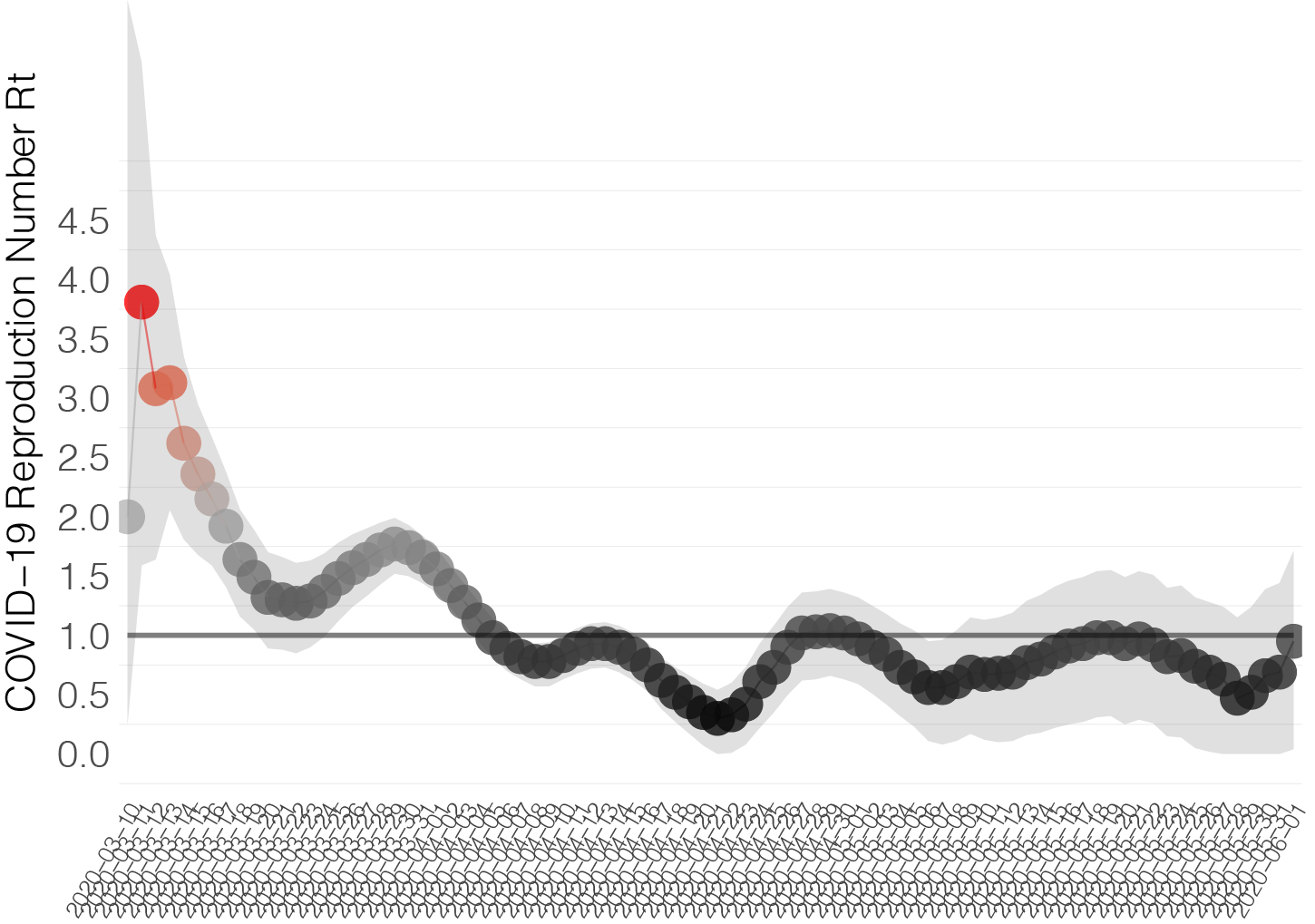}
\caption{Median value (red) and 95$\%$ credible intervals  for the posterior distributions of the effective reproduction number in compartmental model 3, for the period between $05/03/2020$ until $01/06/2020$.}
\label{model3_Rt}
\end{figure}

We then use the methodology proposed  by~\cite{bettencourt2008real} and recently modified by~\cite{systrom2020metric} as described in detail in Section~\ref{sec:compmodel4}. For that method we also use the incidents from local transmission in Cyprus as were reported by the Ministry of Health. Figure~\ref{model3_Rt} shows the daily median value as well as the 95\% credible intervals for the effective reproductive number using that method.

\begin{figure}[H]
\centering\includegraphics[width=0.7\textwidth]{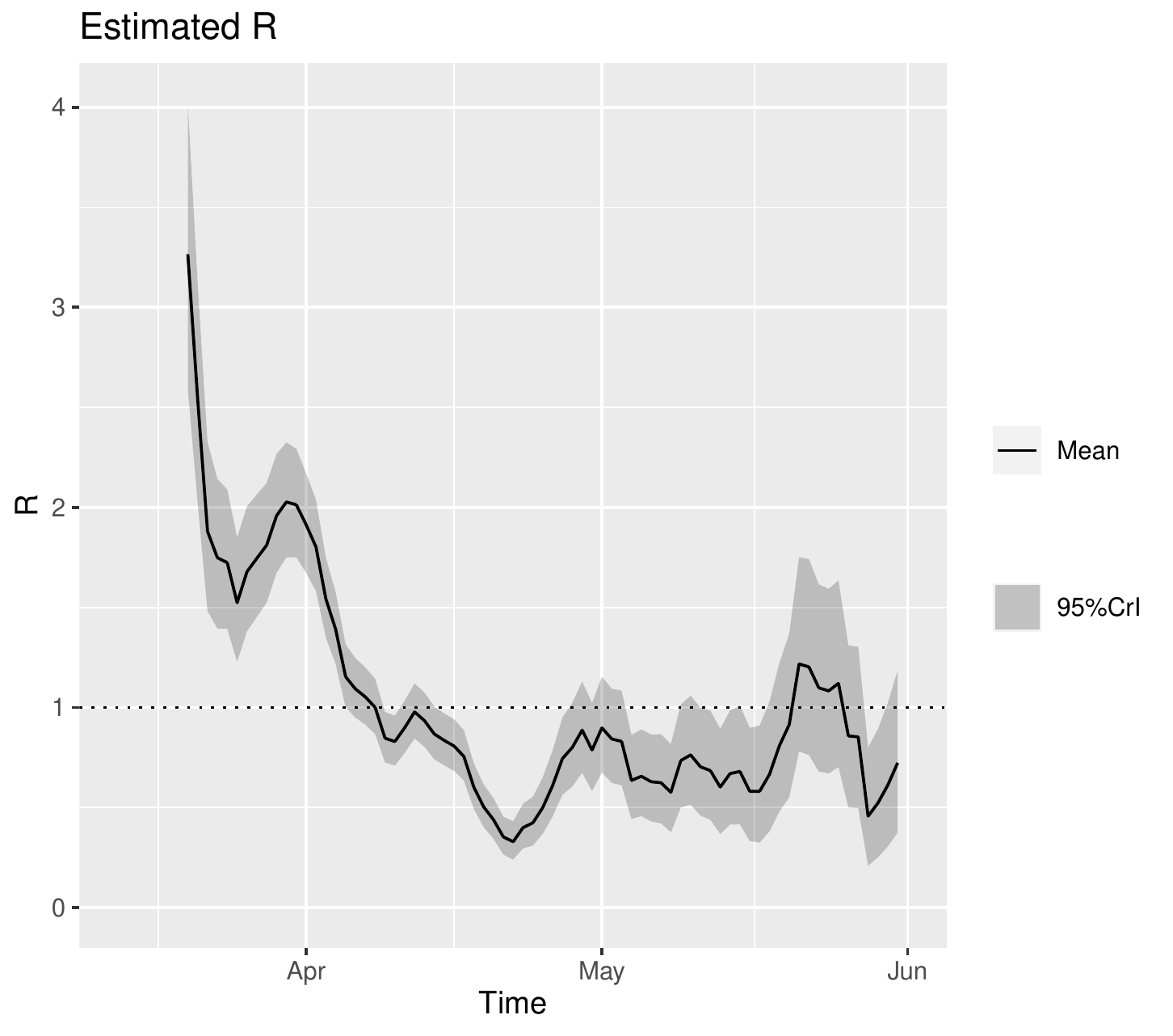}
\caption{Estimation of $R_{t}$ based on the methodology of \citet{Corietal(2013)} together with 95\%
confidence (green curve) intervals. The plot is  based on weekly data and on choosing the serial interval distribution to be Gamma with  mean 6.48 days and standard deviation of 3.83;
see \cite{Fergimperial(2020)} for more.}
\label{fig:CorietalR_estimates}
\end{figure}

These results should be compared with the methodology of \citet{Corietal(2013)}--see Figure \ref{fig:CorietalR_estimates} which
shows weekly estimates of $R_{t}$ based on weekly time intervals. At the end of May, COVID-19 was well contained in Cyprus especially even though the disease initiated with a high value of $R_{t}$. The government lockdown helped reduce the reproduction number, as the data shows. Comparing the results obtained by all methods presented in
this Section, we see no gross discrepancies on concluding that $R_{t} < 1$ by the end of May, with high probability.

\section{Discussion}

The work presented in this report is the result of intensive collaboration of
an interdisciplinary team which was formed shortly after the pandemic started. The main motivation was to give guidance to Cypriot government for controlling this major infectious disease outbreak. Accordingly, we developed models and methods  that
are of  critical importance  in appreciating how this disease  is developing and what will be its next stage  and in what kind of time framework. This is a valuable information  for outbreak control, resource utilization and to initiate again the normal daily life.

We followed diverse paths to accomplish this by appealing to different modeling approaches and methods.  We have shown that the government interventions were successful  on containing  COVID-19 in Cyprus, by the end of May,  even though the disease initiated with a high value of $R_{t}$. The government lockdown helped reduce the reproduction number, as the data shows, by applying different methodology. In addition, we have shown by change-point methodology and time series analysis the effect of various measures taken and have developed short-term predictions.
The models we applied are based on simple surveillance data, seem to work well, give similar results, and can certainly help epidemiologists and public health officials quantify and understand changes in the transmission intensity of future epidemics and the drivers of these changes. Finally, we feel that our approach to bring together experts from various fields avoids misunderstandings and gaps in communication between scientists, and maximizes the effectiveness of efforts to deal with public health emergencies.

\renewcommand{\theequation}{A-\arabic{equation}}
\renewcommand{\theLemma}{A-\arabic{Lemma}}
\setcounter{equation}{0}
\setcounter{Lemma}{0}
\renewcommand{\thesubsection}{A-\arabic{subsection}}
\setcounter{subsection}{0}
\section*{Appendix}

\subsection{Isolate-Detect Methodology}
\label{sec:ISDmethod}
The existing change-point detection techniques for the scenarios mentioned in Section \ref{subsub:cpt_detection_methods} are mainly split into two categories based on whether the change-points are detected all at once or one at a time. The former category mainly includes optimization-based methods, in which the estimated signal is chosen based on its least squares or log-likelihood criterion  penalized by a complexity rule in order to avoid overfitting. The most common example of a penalty function is the Bayesian  Information Criterion (BIC); see \cite{Schwarz(1978)} and \cite{Yao} for details. In the latter category, in which change-points are detected one at a time, a popular method is binary segmentation, which performs an iterative binary splitting of the data on intervals determined by the previously obtained splits. Even though binary segmentation is conceptually simple, it has the disadvantage that at each step of the algorithm, it looks for a single change-point, which leads to its suboptimality in terms of accuracy, especially for signals with frequent change-points. One method that works towards solving this issue is the Isolate-Detect (ID) methodology of \cite{Anastasiou_Fryzlewicz}; it is the method used for the analysis carried out in this paper.


The concept behind ID is simple and is split into two stages; firstly, the isolation of each of the true change-points within subintervals of the domain $[1,2,\ldots,T]$, and secondly their detection. The basic idea is that for an observed data sequence of length $T$ and with $\lambda_T$ a positive constant, ID first creates two ordered sets of $K = \left\lceil T/\lambda_T \right\rceil$ right- and left-expanding intervals as follows. The $j^{th}$ right-expanding interval is $R_j = [1,j{\lambda_T}]$, while the $j^{th}$ left-expanding interval is $L_{j} = [T - j\lambda_T + 1,T]$. These intervals are collected in the ordered set $S_{RL} = \left\lbrace R_1, L_1, R_2, L_2, \ldots,R_K,L_K\right\rbrace$. For a suitably chosen contrast function, ID identifies the point with the maximum contrast value in $R_1$. If its value exceeds a threshold, denoted by $\zeta_T$, then it is taken as a change-point. If not, then the process tests the next interval in $S_{RL}$. Upon detection, the algorithm makes a new start from the end-point (or start-point) of the right- (or left-) expanding interval where the detection occurred.

\begin{figure}[htp]
\centering\includegraphics[width=0.8\textwidth,]{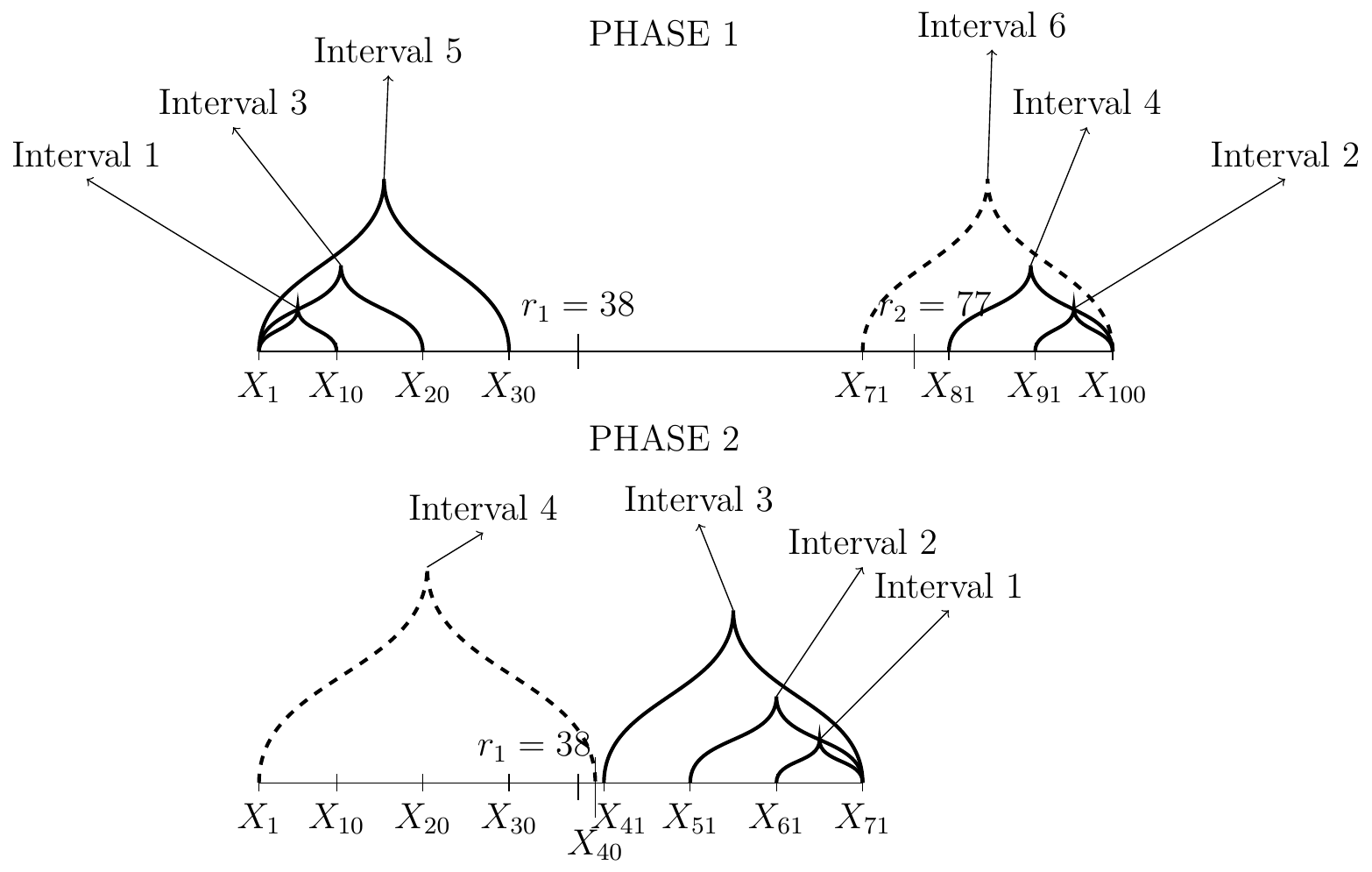}
\caption{An example with two change-points; $r_1=38$ and $r_2 = 77$. The dashed line is the interval in which the detection took place in each phase.}
\label{fig:idea_intro}
\end{figure}

For clarity of exposition, we give below a simple example. Figure \ref{fig:idea_intro} covers a specific case of two change-points, $r_1=38$ and $r_2=77$. We will be referring to Phases 1 and 2 involving six and four intervals, respectively. These are clearly indicated in the plot and they are only related to this specific example, as for cases with more change-points will entertain more  such phases. At the beginning, $s=1$, $e=T=100$, and we take the expansion parameter $\lambda_T = 10$. Then, $r_2$ gets detected in $\left\lbrace X_{s^*}, X_{s^*+1},\ldots,X_{e}\right\rbrace$, where $s^*=71$. After the detection, $e$ is updated as the start-point of the interval where the detection occurred; therefore, $e=71$. In Phase 2 indicated in the plot, ID is applied in $[s,e]=[1,71]$. Intervals 1, 3 and 5 of Phase 1 will not be re-examined in Phase 2 and $r_1$ gets, upon a good choice of $\zeta_T$, detected in $\left\lbrace X_{s}, X_{s+1},\ldots,X_{e^*}\right\rbrace$, where $e^*=40$. After the detection, $s$ is updated as the end-point of the interval where the detection occurred; therefore, $s=40$. Our method is then applied in $[s,e] = [40,71]$; supposing there is no interval $[s^*,e^*]\subseteq [40,71]$ on which the contrast function value exceeds $\zeta_T$, the process will terminate.

\subsection{More on Count Time Series Models and Interventions}
\label{Sec:CountTimeSeries}

Recall \eqref{eq:log-linear model} and that
the parameters $d,a_{1},b_{1}$  can be positive or negative but they need to satisfy certain conditions so that we obtain stable behavior of the process. Note that the lagged
observations of the response $X_t$ are fed into the autoregressive equation for $\nu_t$ via the term $\log(X_{t-1}+1)$.
This is a  one-to-one transformation of $X_{t-1}$ which avoids  zero data values.
Moreover,   both $\lambda_{t}$ and $X_{t}$ are transformed  into  the same scale. Covariates can be  easily accommodated by model \eqref{eq:log-linear model}. When $a_{1}=0$, we obtain an AR(1) type  model in terms of $\log(X_{t-1}+1)$. In addition,
the log-intensity process of \eqref{eq:log-linear model} can be rewritten as
\begin{eqnarray*}
\nu_{t}= d \frac{1-a_{1}^{t}}{1-a_{1}} + a_{1}^{t} \nu_{0} + b_{1} \sum_{i=0}^{t-1} a_{1}^{i} \log(1+X_{t-i-1}),
\end{eqnarray*}
after repeated substitution. Hence, we obtain again that the hidden process $\{\nu_t\}$ is determined by past functions
of lagged responses, i.e. \eqref{eq:log-linear model} belongs to
the class of  observation driven models; see \cite{Cox(1981)}.

Models like \eqref{eq:log-linear model} can accommodate  various type of interventions (or extraordinary) observations
by suitable modification.
Generally speaking,  intervention effects on time series
data are classified according to whether their impact is concentrated on a single or a
few data points, or whether they affect the whole process from some  specific time
$t=\tau$  on. In classical linear time series methodology an
intervention effect is included in the observation equation by
employing a sequence of deterministic covariates $\{ W_t \}$ of the form
\begin{equation}
W_{t} = \xi(\cB)I_t(\tau),
\label{eq:outlier process}
\end{equation}
where $\xi(\cB)$ is a
polynomial operator, $\cB$ is the shift operator such that $\cB^{i} W_t = W_{t-i}$  and  $I_t(\tau)$ is an indicator function,
with  $I_t(\tau)=1$ if $t=\tau$, and $I_t(\tau)=0$ if $t \neq \tau$ .
The choice of the operator $\xi(\cB)$ determines the kind of intervention effect:
additive outlier (AO), transient shift (TS), level shift (LS) and
innovational outlier (IO). Since models of the form \eqref{eq:log-linear model}  are not defined in
terms of innovations, we focus on the first three types of
interventions.  A model like \eqref{eq:log-linear model}, is determined by a latent
process. Therefore  a  formal linear  structure, as in the case of Gaussian linear time series model
does not hold any more and interpretation of the interventions is a more complicated issue.
Hence, a method which allows detection of interventions and estimation of their
size is needed  so that  structural changes can be identified successfully. Important steps to achieve this goal are
the following; see \citet{ChenandLiu(1993)}:
\begin{enumerate}
\item  A  suitable model for accommodating interventions in count time series data.
\item  Derivation of test procedures for their successful detection.
\item  Implementation of joint maximum likelihood estimation of model parameters and outlier sizes.
\item  Correction of the observed series for the detected interventions.
\end{enumerate}
All  these issues and  possible directions for further developments of the methodology have been addressed by \cite{tscountR}
under the Poisson and mixed Poisson distributional framework.

\subsection{Computational Details for Fitting Equations \eqref{eq:model3}}
\label{Sec:competailsModel3}

\begin{figure}[H]
\centering\includegraphics[width=0.40\textwidth,height=0.65\textheight]{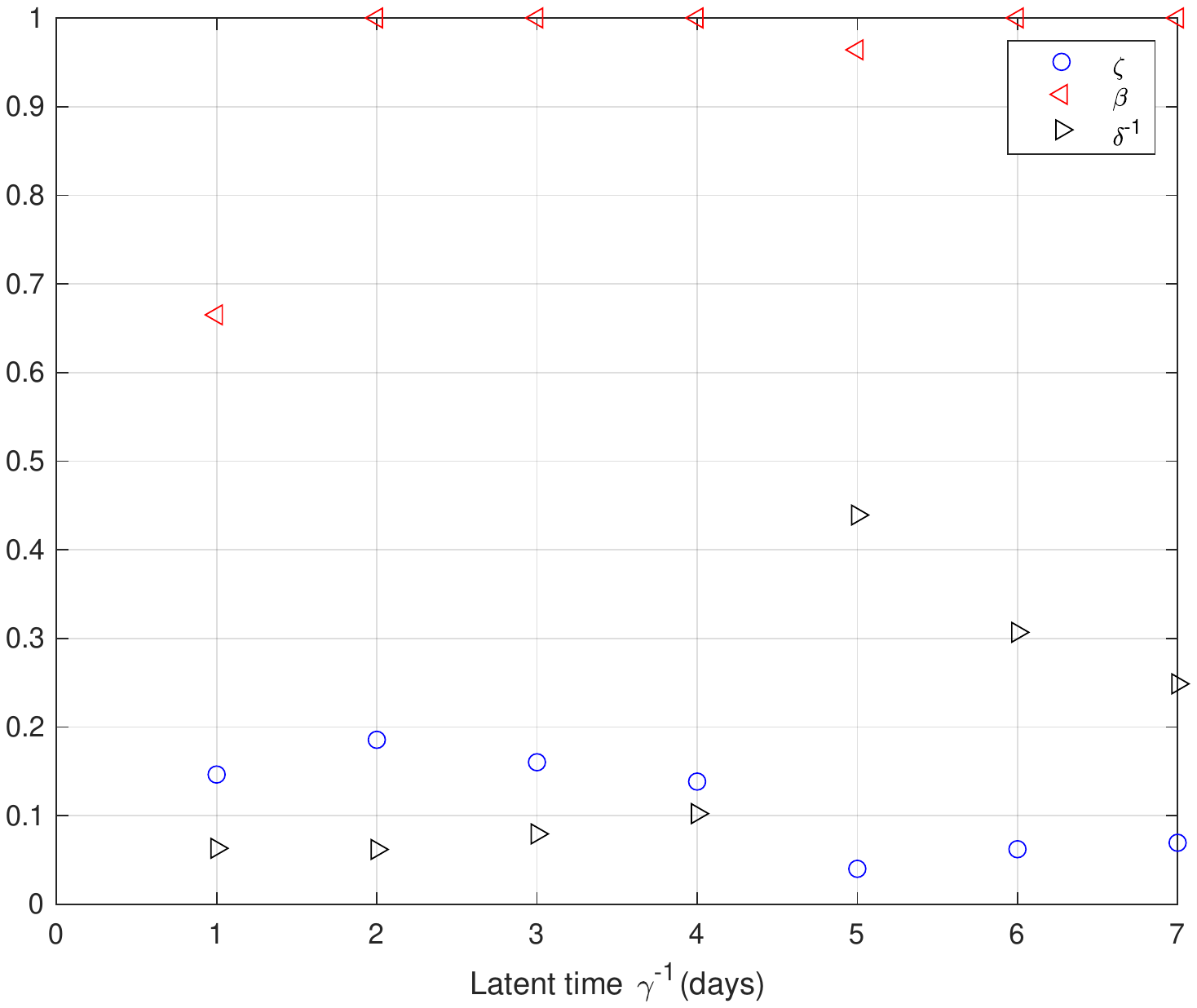}
\centering\includegraphics[width=0.40\textwidth,height=0.65\textheight]{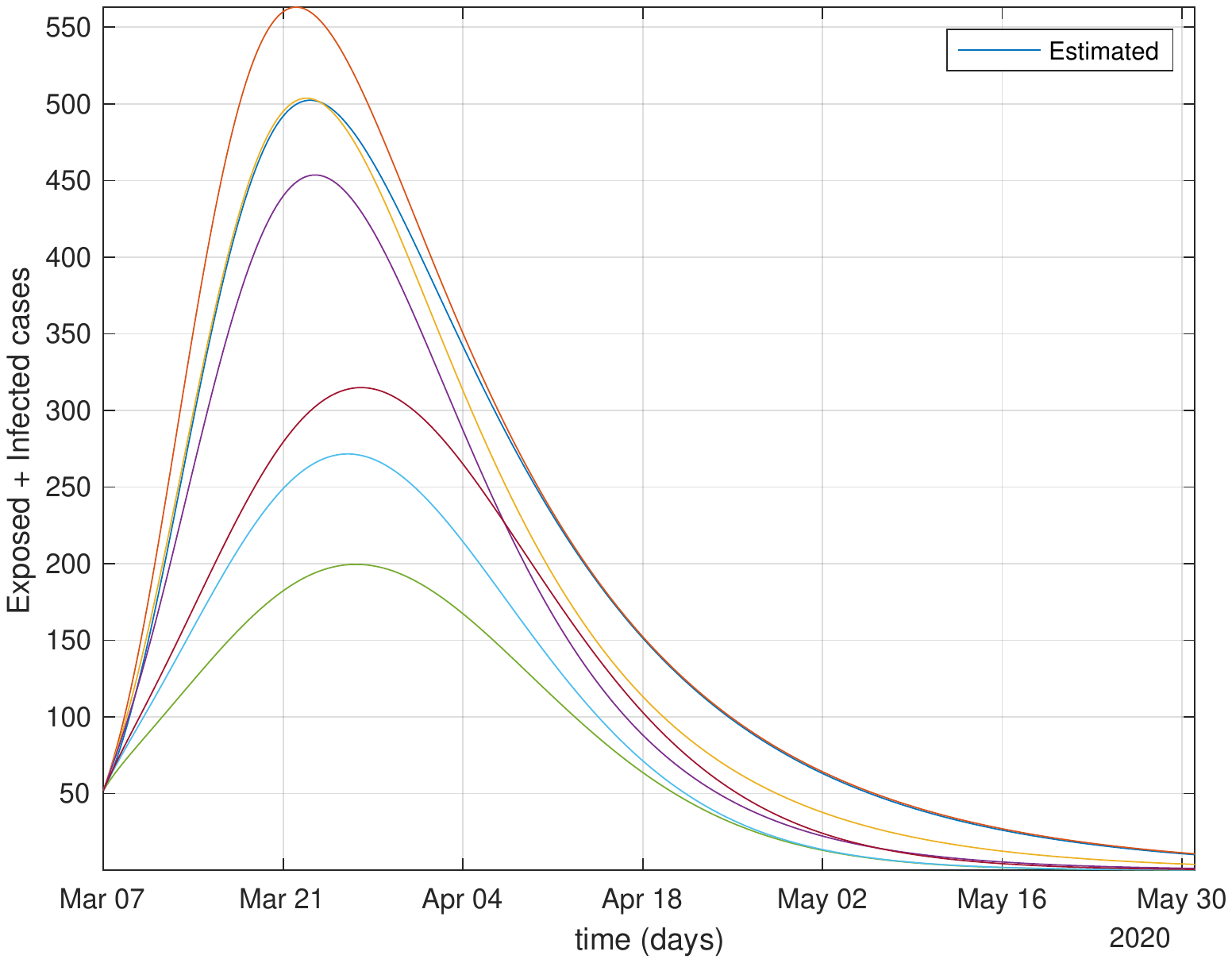}
\vspace{-4cm}
\caption{Sensitivity analysis on the parameters for the model defined by \eqref{eq:model3}.  The influence of the latent time $\gamma^{-1}$ on the protection rate $\zeta$, the transmission rate $\beta$ and the quarantine time $\delta^{-1}$ (left plot ), the sum of exposed and infectious cases $E(t)+I(t)$ (right plot).}
\label{fig:gamma_inf}
\end{figure}

According to the official reports, the number of quarantined cases ($Q$), recovered ($R$) and deaths ($D$), due  to COVID-19, are available. However, the recovered and death cases are directly related to the number of quarantine cases, which plays an important role in the analysis, especially since the numbers of exposed ($E$) and infectious ($I$) cases are very hard to determine. The latter two are therefore treated as hidden variables. This implies that we  need to  estimate  the four parameters $\zeta,\beta, \gamma^{-1}, \delta^{-1}$  and both the time dependent cure rate $\lambda(t)$ and mortality rate  $\kappa(t)$. This is an optimization problem that we solve  as follows: first we allow the latent time $\gamma^{-1}$ to vary between 1 and 7 days and for each fixed $\gamma^{-1}$,  we explore its influence on the rest of the parameters. The system of differential  equations \eqref{eq:model3} is solved numerically using the Runge-Kutta 45 numerical scheme. 
The left plot of  Figure \ref{fig:gamma_inf}  shows  that the protection rate $\zeta$  and the transmission rate $\beta$ both attain their corresponding maximum value when  $\gamma^{-1}$ is  equal to 3 days. Note
that $\zeta$ takes values between 0.08 and 0.2, while $\beta$ converges very fast to 1. The reciprocal of the quarantine time $\delta^{-1}$  is increasing with the latent time $\gamma^{-1}$. One would suspect  that longer latent time results to higher transmission rate and as the latent time increases almost every unprotected person will be infected after a direct contact with a COVID-19 patient. The right plot of  Figure \ref{fig:gamma_inf} shows  the effect of the latent time on the total number of infected cases (exposed and infectious $E(t)+I(t)$) but not yet quarantined. The peak of the infection was achieved between the 21st and the 24th  of March, depending on the latent time with the estimated number of infected people ranging between 338 and 526, depending again on the  latent time considered. Hence, once the latent time $\gamma^{-1}$ is fixed, the fitting performance depends on the values of $\zeta, \beta$ and $\delta^{-1}$.  After a small sensitivity analysis the latent time was finally determined as 3 days. The mortality rate $\kappa(t)$  is constantly very small and almost equal to zero, therefore we have not attempted  to fit any function to it. For  the cure rate $\lambda(t)$ we have fitted the exponential function given in \ref{eq:recovery},
the idea behind being that with time the recovery should converge to a constant rate.  For the parameter estimation we have used a modified version of the MATLAB code given by \cite{Cheynet2020} because Cyprus
is a small country and this fact needs to be taken properly into account.

\subsection{Prior Modelling and Computational Details for Sec. \ref{sssec:mod12}}
\label{Sec: CompdetailsforR}
We present a Bayesian analysis, for the model defined by \eqref{eq:Compmodel1} in Sec. \ref{sec:compmodel1},
for the parameters  $\alpha$ and $\beta$, based on  six separate fortnight periods. We use Beta prior distributions on the reporting rate $\alpha\in[0,1]$, with parameters reflecting the amount of targeted and random testing performed in Cyprus, at the time under consideration. In particular, in the first period (when the number of tests was relatively low) we employ a symmetric prior around the value $\alpha=0.5$, while for later periods (when the number of targeted and random tests increased) we let the prior become progressively skewed towards 1. For the transmission rate $\beta>0$, in the first period we use a $Gamma(3/2,3/2)$ prior, which puts high probability around 2, while for later periods we use an $Exponential(1)$ prior which puts more mass closer to zero. This choice reflects the existence of super-spreaders in the early stages of the outbreak with higher probability compared to later on.

In each time-period under consideration we also need to initialize the outbreak in Cyprus. For the first period in both datasets, we use a uniform prior supported in $\{0,1,\dots,10\}$ on the number of exposed and the number of undocumented infected 3 days before the first recorded incident. The two priors are independent, while the number of susceptible individuals is taken equal to Cyprus' population and the number of infected-reported equal to zero.  For later periods, we use as priors on the four state variables, their posterior distributions at the end of the previous period (corrected appropriately based on the observation at the end of the previous period).

Following \cite{Li2020}, we assume that the daily number of reported cases are independent Gaussian random variables and use an empirical variance given as
$$
\sigma^2(t)=\max\Bigl(1, \frac{y(t)^2}4\Bigr),
$$
by recalling that $y(t)$ denotes the number of infected cases at day $t$.
This allows us to build a Gaussian likelihood for the parameters $\alpha$ and $\beta$. Combining this likelihood with the prior distributions, we can deduce a formula for the posterior distribution on $\alpha, \beta$. This distribution is not available in a closed form, hence in order to compute posterior estimates and their respective uncertainty quantification, we need to sample it. In the relatively simple setting of Model 1, it is feasible to employ Markov chain Monte Carlo methods, (see \cite{RC13}),
in order to sample the posterior (namely, we use an independence sampler). This is in contrast to the model defined by \eqref{eq:Compmodel2} in Sec. \ref{sec:compmodel2}, see \cite{Li2020}, where one has to use the  Ensemble Adjustment Kalman Filter (EAKF), which introduces some approximations to the posterior distribution, due to the more complex meta-population structure.
Originally developed for use in weather prediction, the EAKF assumes a Gaussian distribution for both the prior and the likelihood and adjusts the prior distribution to a posterior using Bayes rule deterministically. In particular, the EAKF assumes that both the prior distribution and likelihood are Gaussian, and thus can be fully characterized by their first two moments (mean and variance). The update scheme for ensemble members is computed using Bayes rule (posterior $\sim$ prior $\times$ likelihood) via the convolution of the two Gaussian distributions (see \cite{Li2020} for the implementation).


\subsection{Results of Change-Point Analysis for Piecewise-Constant Model}
\label{Sec: Piecewise Constant}

We report the results obtained after fitting  a piecewise-constant signal plus noise model, as descibed in Sec. \ref{subsub:cpt_detection_methods}.
The scenario here is that at each change-point, we have a sudden jump in the mean level of the signal. Figure \ref{sampling_cases_constant} below gives a graphical representation of the relevant change-point analysis carried out for the daily number of  COVID-19 cases.
Based on this piecewise-constant scenario, the Isolate-Detect methodology used has detected five important changes leading to six homogeneous periods in terms of the average number of detected cases per day. Changes are detected on the $11^{th}$ and the $25^{th}$ of March, on the $2^{nd}$ and the $14^{th}$ of April, as well as on the $1^{st}$ of May.
The first change-point on the $11^{th}$ of March introduces a jump of magnitude 10.56 up to two decimal places, meaning that in the second period there is a mean growth of about 11 detected cases compared to the first period. The initial period mainly consists of days without any new cases detected and therefore, the first change-point is related to the outbreak of the epidemic in the society. The second change-point on the $25^{th}$ of March indicates an elevation of the number of cases by 23.21, which is very important and captures the development of the clusters A and B as mentioned in Section \ref{ssec:surveillance}. The third and fourth change-points, detected on the $2^{nd}$ and $14^{th}$ of April show a reduction in the number of cases, with the number falling by 9.67 and 17.24, respectively. Both these change-points show the importance (in fighting the virus) of the Government's decrees on the $24^{th}$ and $31^{st}$ of March for a general lockdown. The last change-point indicates a further reduction in the number of cases with magnitude 4.95. Corresponding predictions obtained by this model show that there will be no more than 7 new cases per day with the point estimator being equal to 3 cases per day. Our analysis under piecewise-constancy gives two more change-points than those detected when the piecewise-linear structure was employed in Section \ref{subsub:cpt_detection_application}; in fact, it is always expected to detect more change-points under piecewise-constancy than under piecewise-linearity. For example, think of a noiseless linear signal with upward trend that does not have any change-points in the first derivative. Treating this signal under the piecewise-constant scenario, would mean that each data point is a change-point because a jump of magnitude equal to the slope of the signal is introduced at every time point.

\begin{figure}[H]
\centering\includegraphics[width=0.8\textwidth,height=0.30\textheight]{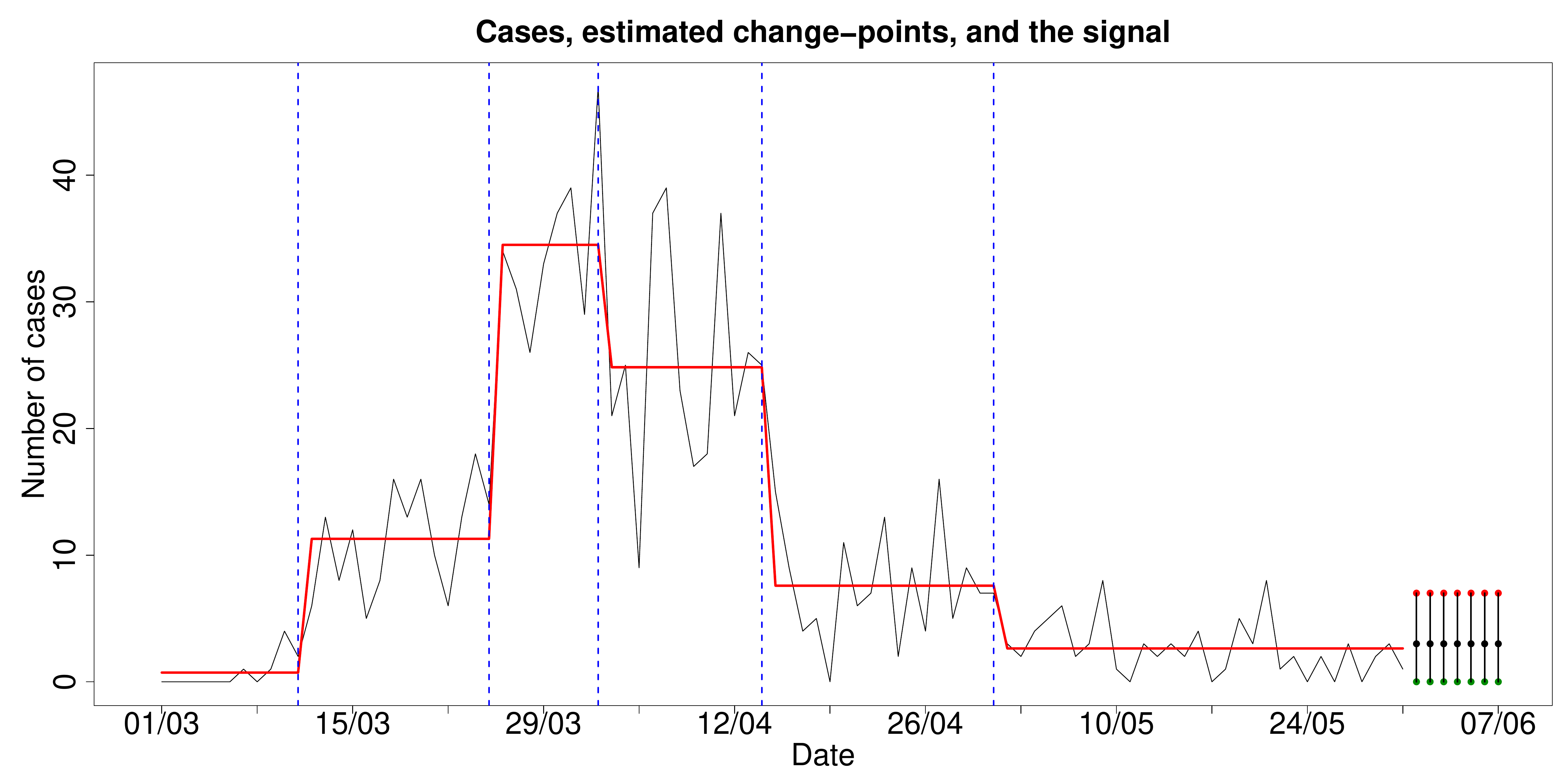}
\caption{The real data (black coloured line) and the estimated piecewise-linear fit (red coloured line) for the daily incidence rate. The change-point locations are given with dotted, blue vertical lines. At the right-end of the plot  point estimators (black dots) and $95\%$ prediction intervals for the number of daily cases for the next week  are reported.}
\label{sampling_cases_constant}
\end{figure}

\subsection{Additional Figures}
\label{Sec:AddFigures}

\begin{figure}[H]
\centering
\includegraphics[width=0.8\textwidth]{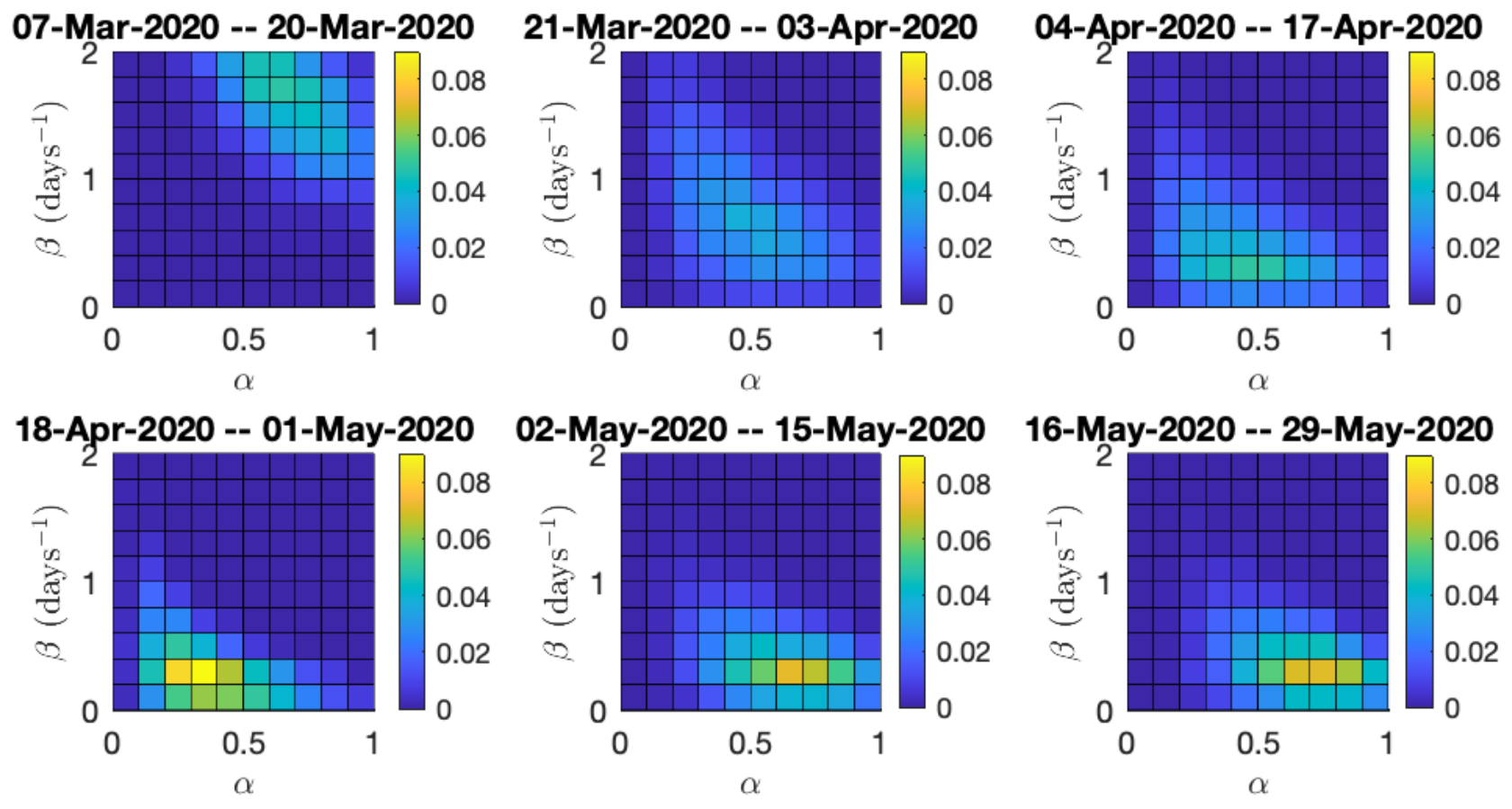}
\caption{Joint posterior distributions of the reporting rate $\alpha$ and the transmission rate $\beta$ in Model 1, for the six fortnight periods starting from $07/03/2020$ until $29/05/2020$. Analysis using data on local transmission only.}
\label{ab_post_local}
\end{figure}
\begin{figure}[H]
\centering\includegraphics[width=0.8\textwidth]{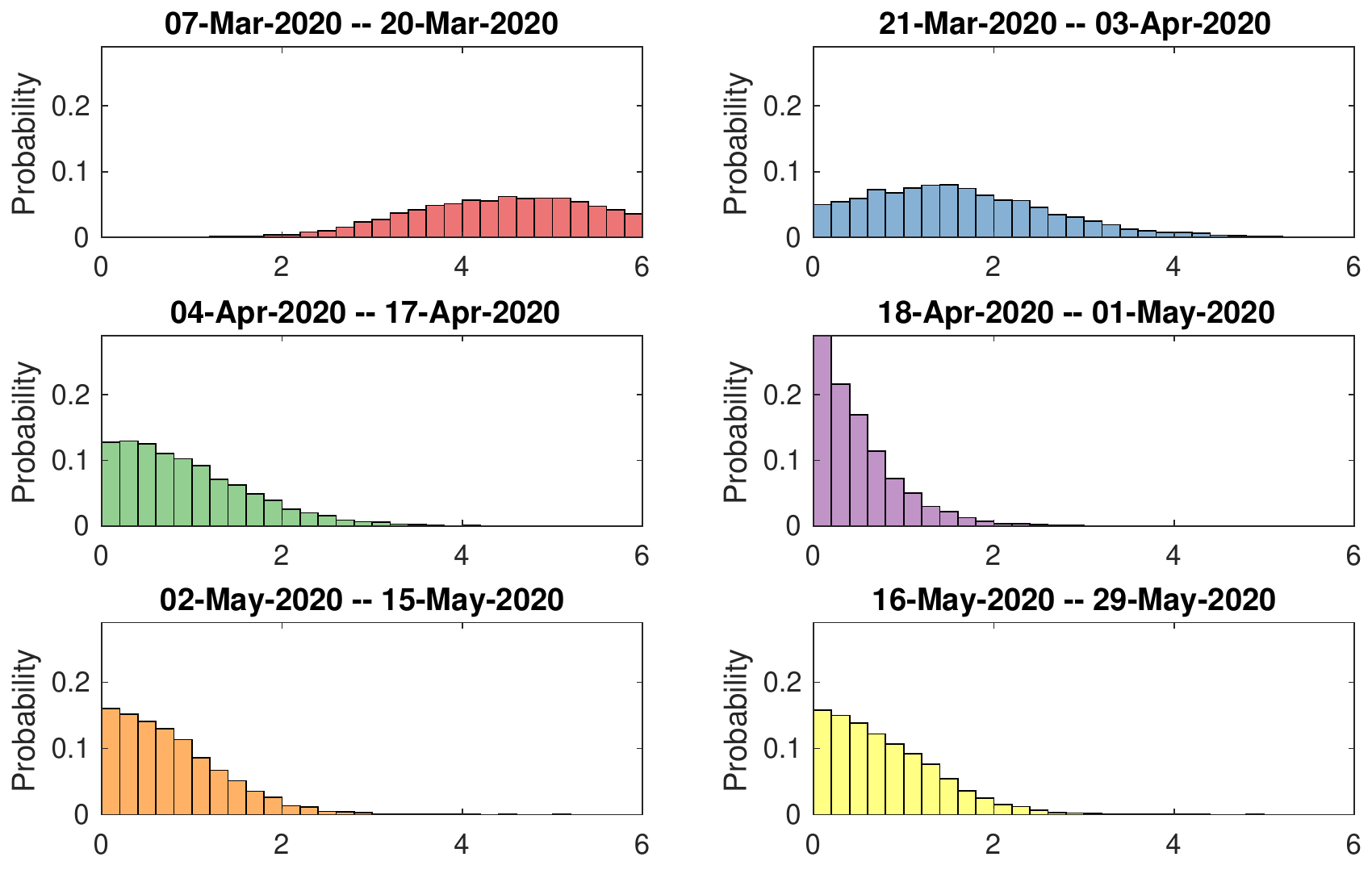}
\caption{Posterior distributions of the effective reproductive number in Model 1, for the six fortnight periods starting from $07/03/2020$ until $29/05/2020$. Analysis using data on local transmission only.}
\label{R_hists_local}
\end{figure}

\begin{figure}[H]
\centering\includegraphics[width=0.7\textwidth]{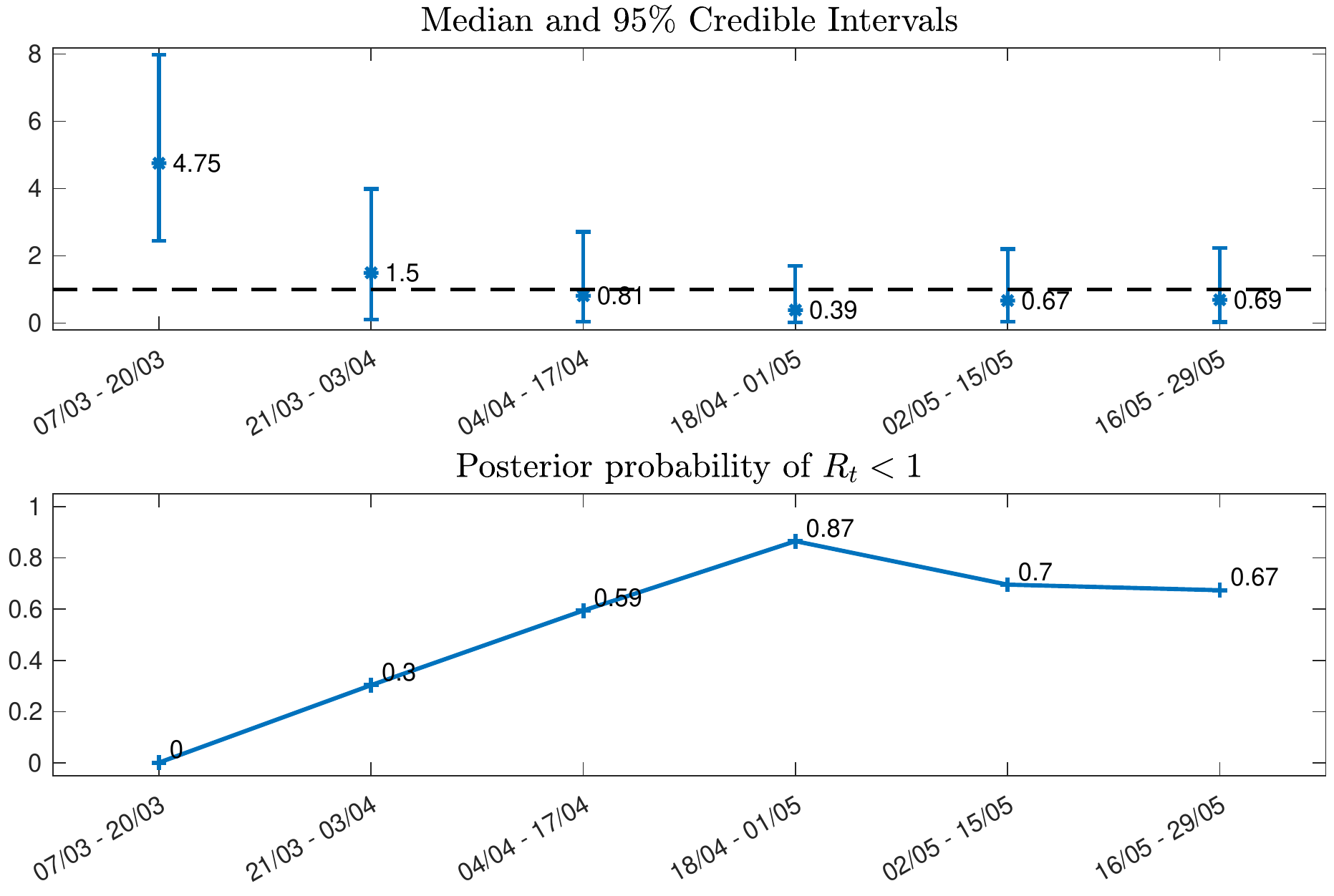}
\caption{Median and 95$\%$ credible intervals for the posterior distributions of the effective reproduction number in Model 1, for the six fortnight periods starting from $07/03/2020$ until $29/05/2020$ (top). Posterior probabilities of the event $R_t<1$ (bottom). Analysis using data on local transmission only.}
\label{R_estimates_local}
\end{figure}

\subsection*{Data availability} Data and code are available at GitHub (\url{https://github.com/chrisnic12/covid_cyprus})

\clearpage

\small{
\bibliography{Covidbib}
}

\end{document}